\documentclass{PoS}

\def\msbar{\overline{\mathrm{MS}}}

\newcommand{\matel}[3]{\langle #1|#2|#3\rangle}

\newcommand{\ym}{y_m}

\title{The conformal window on the lattice}

\ShortTitle{Conformal window}

\author{\speaker{Luigi Del Debbio}\thanks{LDD is supported by an STFC Advanced Fellowship.}\\
        School of Physics and Astronomy, University of Edinburgh\\
        Edinburgh EH9 3JZ, United Kingdom \\
        E-mail: \email{luigi.del.debbio@ed.ac.uk}}

      \abstract{Lattice simulations can play an important role in the
        study of dynamical electroweak symmetry breaking by providing
        quantitative results on the nonperturbative dynamics of
        candidate theories. For this programme to succeed, it is
        crucial to identify the questions that are relevant for
        phenomenology, and develop the tools that will provide robust
        answers to these questions. The existence of a conformal
        window for nonsupersymmetric gauge theories, and its
        characterization, is one of the phenomenologically important
        problems that can be studied on the lattice. We summarize the
        recent results from studies of IR fixed points by numerical
        simulations, discuss their current limitations, and analyze
        the future perspectives. }

\FullConference{The XXVIII International Symposium on Lattice Field Theory, Lattice2010\\
		June 14-19, 2010\\
		Villasimius, Italy}

\begin{document}

\section{Motivations and aims}
\label{sec:mot}

\begin{figure}[ht]
  \centering
  \includegraphics[width=3.0truein]{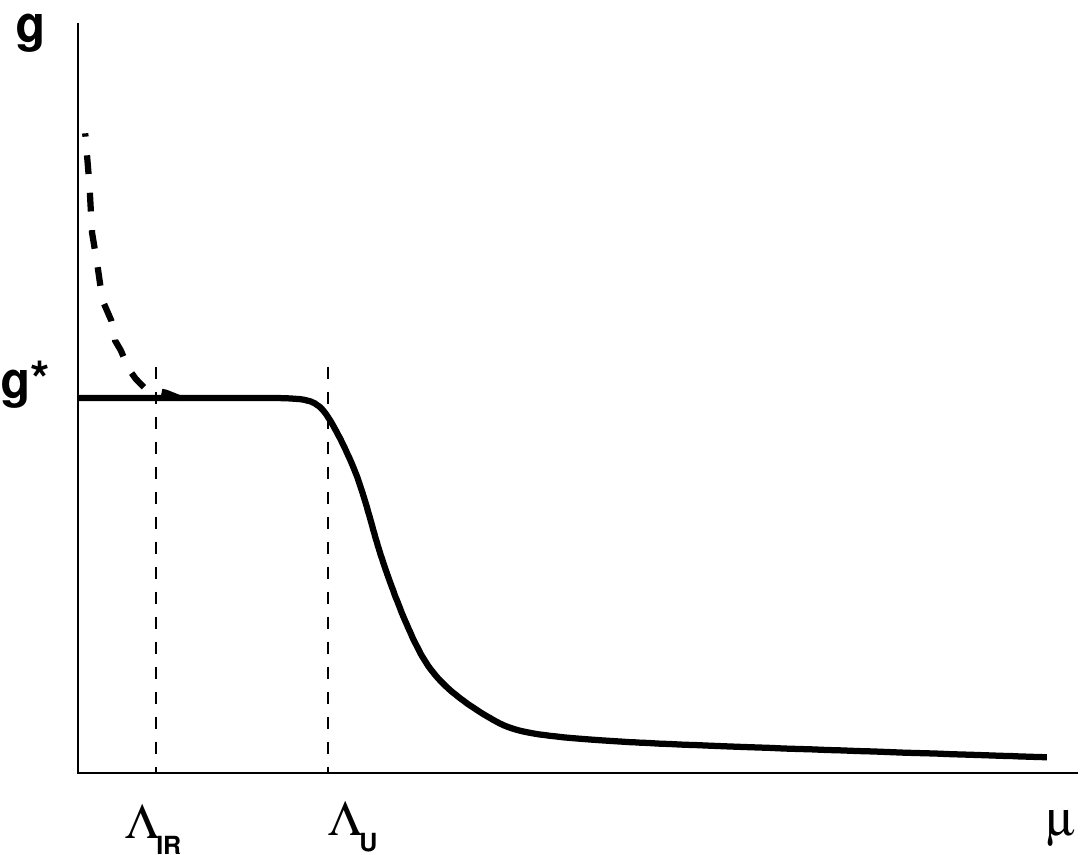}
  \caption{Running of the coupling as a function of the energy scale
    for a theory with an IRFP. At low energies the coupling flows to a
    fixed-point value $g^*$, while the high energy behaviour is the
    usual one expected for asymptotically free theories. The scale
    $\Lambda_\mathrm{U}$ corresponds approximately to the energy where
    the running starts to be dictated by asymptotic freedom. The
    dashed curve at low energies shows the running of the coupling
    when a fermionic mass term is switched on.  }
  \label{fig:beta}
\end{figure}

Theories that are asymptotically free at high energies, and have an
infrared fixed point (IRFP) in the renormalization group (RG) flow of
their couplings are said to be inside the ``conformal window''. The
typical running of the coupling, with its limiting value $g^*$ at
small energies, is shown in Fig.~\ref{fig:beta}. The existence of a
conformal window has been studied analytically for supersymmetric
theories, see e.g. Ref.~\cite{Intriligator:1995au} for a
review. Quantitative analytical studies of the IR regime for the
non-supersymmetric cases are more difficult, due to a lack of adequate
tools to investigate the nonperturbative dynamics of gauge
theories. Nevertheless interesting results have appeared in recent
years, see e.g.\
Refs.~\cite{Sannino:2009za,Poppitz:2009uq,Nunez:2008wi,Armoni:2009jn}
for a summary. Besides its field-theoretical interest, the existence
of a conformal window has important consequences for building models
of physics beyond the standard model~(BSM) that are based on
strongly-interacting theories.

Models of dynamical electroweak symmetry breaking~(DEWSB) provide
elegant realizations of the mechanism responsible for the breaking of
electroweak symmetry down to electromagnetism, by means of the vacuum
expectation value of a fermion bilinear, see
Ref.~\cite{Nakamura:2010zzi} for a review and a list of references. In
the simplest versions of DEWSB, the fermion condensate is produced by
the strong dynamics of some new asymptotically free non-abelian gauge
theory. The new gauge group is usually called ``technicolor''
(TC)~\cite{Weinberg:1975gm,Susskind:1978ms}. The massless fermions
coupled to this new gauge field are called ``technifermions''; they
transform under some representation of the new gauge group, and their
left-handed components are weak doublets, so that their condensate
breaks electroweak symmetry in the standard model~(SM). The
technifermion condensate also breaks the global chiral symmetry of the
technicolor theory, in analogy with chiral symmetry breaking in QCD,
and therefore implies the existence of massless ``technipions''. The
number and the dynamics of technipions depend on the details of the
technicolor model. Three of these technipions are ``Higgsed'' and
become the longitudinal components of the $W$ and $Z$ bosons. As a
result, the latter acquire a mass proportional to the technipion decay
constant, and therefore $F_T\sim v \sim 250~\mathrm{GeV}$, where $v$
denotes the Higgs condensate in the SM. This constraint sets the
typical scale of the technihadronic spectrum,
$\Lambda_\mathrm{TC}$. The rest of the technihadron spectrum depends
on the specific theory that is selected to mediate the new strong
force. Being able to compute in the strongly-interacting regime of the
technicolor theories is the main ingredient to understand this
mechanism.

Lattice simulations are a prime tool to investigate nonperturbative
physics from first principles. However numerical studies will play an
important role in BSM studies only if we are able to identify the
questions that are relevant in order to achieve a quantitative
understanding of DEWSB. In this introduction we briefly review some of
the problems that technicolor model building is confronted to, and
analyze the possibility to use numerical results to make progress in
this area.

\paragraph{Constraints on technicolor.}

\begin{figure}[ht]
  \centering
  \includegraphics[width=4.0truein]{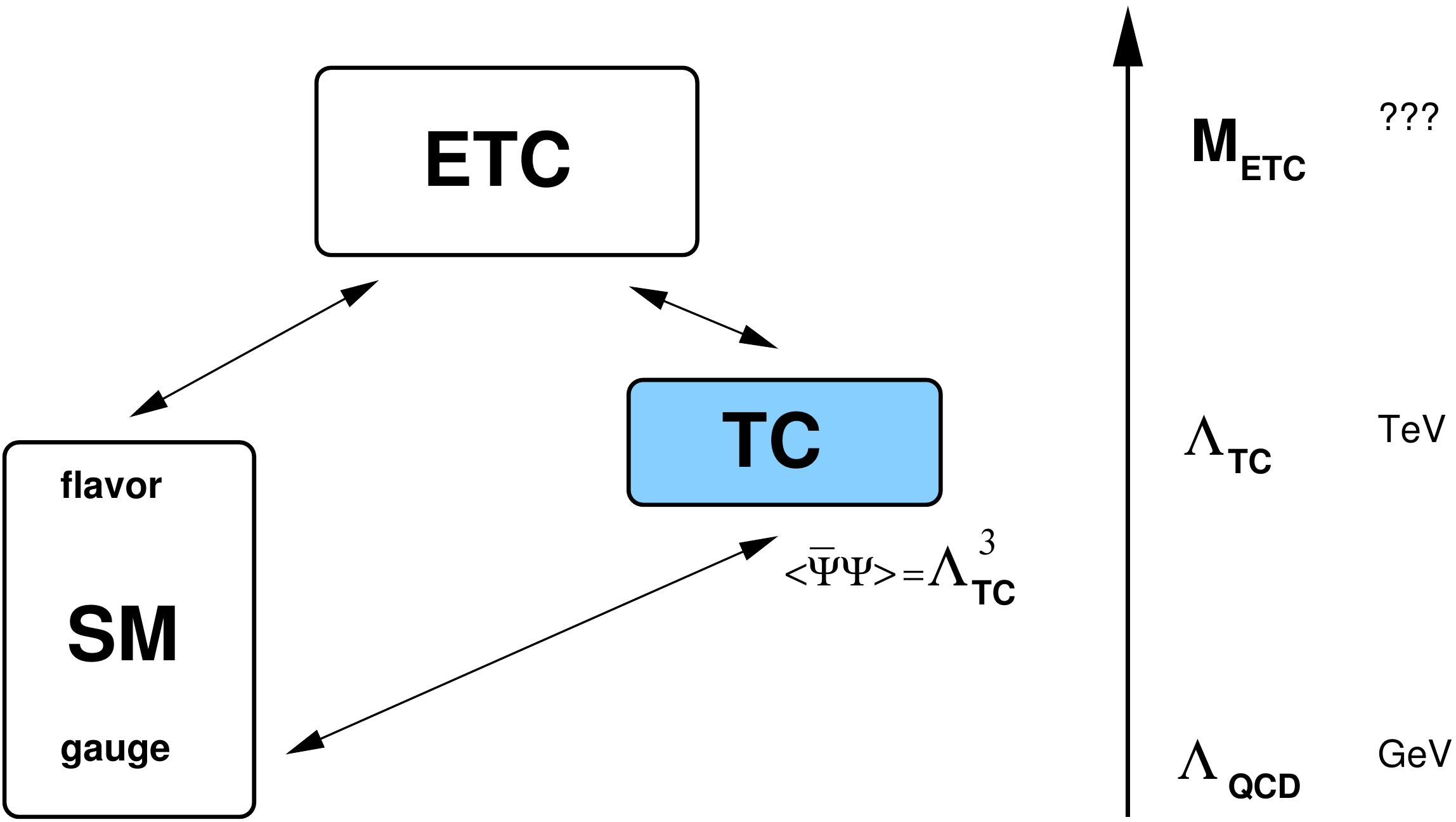}
  \caption{A schematic representation of the technicolor (TC) and
    extended technicolor (ETC) theories, and their connections with
    the standard model (SM). The interactions are characterized by
    different scales, indicated on the right. Bounds on the ETC scale
    $M_\mathrm{ETC}$ are set by SM precision measurements. }
  \label{fig:sketch}
\end{figure}

While technicolor provides a natural way to generate $W$ and $Z$
masses, additional interactions must be introduced in order to
describe the SM flavor sector. A scenario that has been extensively
studied involves ``extended technicolor'' gauge interactions
(ETC)~\cite{Eichten:1979ah,Dimopoulos:1979es}. The ETC gauge bosons
couple to both the SM fermions and the technifermions. At some energy
$M_\mathrm{ETC}$, larger than the TC scale, the extended gauge
symmetry is then broken down to the residual TC gauge symmetry, which
remains intact. As a result, interactions between the technifermions
and the ordinary standard model fermions are generated at
low-energies. We have so far introduced two new sectors, a TC one and
an ETC one. Their connections and the energy scales involved are
summarized in Fig.~\ref{fig:sketch}. Note that below the ETC scale the
TC theory and the flavor sector of the SM are decoupled.

In current simulations, only the TC sector is simulated on the
lattice.  At energies of the order of the TC scale, the ETC-mediated
interactions are described by the effective lagrangian obtained when
the heavy ETC degrees of freedom are integrated out. This lagrangian
includes four-fermion operators, whose couplings are suppressed by
inverse powers of the ETC scale. These are dimension-six operators
involving technifermions and SM fermions; denoting the SM and TC
fermion fields by $\psi(x)$ and $Q(x)$ respectively, the terms that
are relevant for our discussion can be schematically written as:
\begin{eqnarray}
  \label{eq:fourfermi1}
  &&\frac{1}{M_\mathrm{ETC}^2} \bar Q(x) Q(x) \bar \psi(x) \psi(x)\, , \\
  \label{eq:fourfermi2}
  &&\frac{1}{M_\mathrm{ETC}^2} \bar \psi(x) \psi(x) \bar \psi(x)
  \psi(x)\, , 
\end{eqnarray}
where for simplicity we have suppressed the flavor, color and spin
indices.

Eq.~(\ref{eq:fourfermi1}) shows that a technifermion condensate
$\langle \bar Q Q \rangle$ generates a mass term for the SM
fermions. It is worthwhile to recall here that the quark masses in the
SM are defined in a given renormalization scheme and at a given scale
$\mu$, see e.g.\ the review in Ref.~\cite{Nakamura:2010zzi}. When the
fermion masses are generated via an ETC theory, this dependence is
reflected in the scheme and scale dependence of the technifermion
condensate. We are going to follow the convention used in the PDG and
always refer to the quark masses in the $\msbar$ scheme. However it
should be clear that this choice is arbitrary, and that it is possible
to convert from one scheme to the other. Below the scale
$M_\mathrm{ETC}$ the technicolor and the flavor sectors decouple, and
therefore the running of the quark masses is simply given by the QCD
logarithmic evolution, which we neglect here. In an ETC scenario, the
SM fermion masses are therefore set by the value of the technifermion
condensate in the $\msbar$ scheme at the scale $M_\mathrm{ETC}$:
\begin{equation}
  \label{eq:quarkmass}
  m(M_\mathrm{ETC}) = \frac{1}{_{\Lambda^2_\mathrm{ETC}}}
  \left. \langle \bar Q Q\rangle \right|_{M_\mathrm{ETC}} 
  = \frac{1}{_{\Lambda^2_\mathrm{ETC}}} \exp\left[\int_{\Lambda_\mathrm{TC}}^{M_\mathrm{ETC}} 
    \frac{d\mu}{\mu} \gamma(\mu) \right] \left. \langle \bar Q
    Q\rangle \right|_{\Lambda_\mathrm{TC}}\, ,
\end{equation}
where $\gamma$ is the mass anomalous dimension of the technicolor
theory in this scheme, and the second equality shows explicitly the
evolution of the chiral condensate.

On the other hand, ETC interactions in Eq.~(\ref{eq:fourfermi2})
include flavor changing neutral currents (FCNC), which would result in
sizeable contributions to $K$ and $D$ mixing. As noted long
ago~\cite{Eichten:1979ah}, the experimental limits on such
contributions imply a lower bound on the ETC scale. For instance the
ETC scale associated with the generation of the strange mass should be
larger than $10^3~\mathrm{TeV}$, unless some fine-tuned mechanism is
in place. As a consequence of this large value of $M_\mathrm{ETC}$,
heavy quark masses can be generated only if the techniquark condensate
is enhanced with respect to the naive value obtained by rescaling
QCD. A detailed discussion of the fermion mass generation in ETC
models can be found in Ref.~\cite{Appelquist:2003hn}.  Recent
summaries on technicolor and the constraints imposed by current
measurements can be found in
Refs.~\cite{Sannino:2009za,Contino:2010rs,Hill:2002ap}.

It has been suggested that ``walking'' theories, i.e.\ theories where
the RG evolution is very slow, could produce the required enhancement
of the
condensate~\cite{Holdom:1981rm,Holdom:1984sk,Yamawaki:1985zg,Akiba:1985rr,Appelquist:1986an,Appelquist:1987fc,Appelquist:1998xf}. This
is illustrated in Eq.~(\ref{eq:quarkmass}): if the couplings are
approximately constant as the energy scale varies, the integral of the
anomalous dimension yields a power enhancement:
\begin{equation}
  \label{eq:qbarqenh}
  \left. \langle \bar Q Q\rangle \right|_{M_\mathrm{ETC}} 
  = \left(\frac{M_\mathrm{ETC}} {\Lambda_\mathrm{TC}}\right)^\gamma
    \left. \langle \bar Q Q\rangle \right|_{\Lambda_\mathrm{TC}}\, .
\end{equation}
For comparison note that, in a QCD-like theory, the condensate at the
higher scale is of the order of $\left\langle \bar Q Q\rangle
\right|_{\Lambda_\mathrm{TC}}$ up to logarithmic corrections.  For a
sufficiently large value of $\gamma$, the factor in
Eq.~(\ref{eq:qbarqenh}) can yield large quark masses and hence ease
the tension between technicolor and the flavor sector of the
SM. Recent analyses suggest that $\gamma>1$ is needed to avoid a
fine-tuned ETC sector~\cite{Chivukula:2010tn}, while unitarity
constraints imply that $\gamma<2$~\cite{Mack:1975je}. For a
strongly-coupled theory, the value of $\gamma$ can only be computed by
nonperturbative methods.

Furthermore ``walking'' theories are expected to ease the tension
between TC and the constraints from electroweak precision
measurements~\cite{Peskin:1990zt,Altarelli:1990zd,Skiba:2010xn}.

A lucid list of questions that need to be addressed in order to build
a phenomenologically successful model of DEWSB is presented in
Ref.~\cite{Piai:2010ma}.

It is important to recall that the rate at which the coupling
constants evolve, and the corresponding power enhancement, depend on
the renormalization scheme, and therefore the definition of
``walking'' needs to be qualified better. The existence of an IRFP,
i.e.\ a zero of the beta functions for the couplings at small energy
scales, is independent of the scheme. It describes the physical
properties of theories that are scale-invariant at large distances,
where the field correlators have power-like behaviours characterized
by the anomalous dimensions of the fields~\footnote{Anomalous
  dimensions at the fixed point are closely related to the critical
  exponents that are introduced to study statistical systems near
  criticality.}. A concise description of the scheme dependent
features of IRFPs is presented in Ref.~\cite{Bursa:2009we}.

\paragraph{The conformal window.}

In the absence of fermionic matter fields, SU(N) gauge theories are
asymptotically free. The running of the gauge coupling $g$ is encoded
in the beta function, which can be computed in perturbation theory
close to the Gaussian fixed point $g=0$:
\begin{equation}
  \label{eq:betafn}
  \mu\frac{d}{d\mu} g = \beta(g) = -\beta_0 g^3 - \beta_1 g^5 + \ldots\, .
\end{equation}
At one-loop in perturbation theory the effect of the fermion fields can
be read from the first coefficient
\begin{equation}
  \label{eq:beta0}
  \beta_0 = \frac{1}{(4\pi)^2}\left[
    \frac{11}{3} C_2(A) - \frac43 T_R n_f
  \right]\, ,
\end{equation}
where $C_2(R)$ and $T_R$ denote respectively the quadratic Casimir and
the normalization of the generators in the representation $R$. The
coefficient $C_2(A)$ originates from the gluons being in the adjoint
representation of the gauge group, and $n_f$ counts the number of
Dirac fermions in the theory. At one loop the dependence on the
fermionic representation is entirely encoded in the factor $T_R$. In
discussing perturbative results, we use here the same conventions
introduced in Ref.~\cite{DelDebbio:2008wb}. As the number of fermion
fields is increased, $\beta_0$ changes sign and asymptotic freedom is
lost. We shall denote by $n_{f,\mathrm{up}}$ the number of fermions
above which asymptotic freedom is lost, clearly such a number depends
on the gauge group and the fermionic representation.

At higher orders in perturbation theory the fermionic contribution to
the running of the coupling has the potential to generate a
non-trivial zero of the beta function before asymptotic freedom is
lost. This is signalled in perturbation theory by a change of sign of
the coefficient $\beta_1$. We shall denote by $n_{f,\mathrm{lo}}$ the
number of fermions above which the theory exhibits an IRFP. In this
case the theory becomes scale-invariant at large distances, while the
short-distance behaviour is still the one dictated by asymptotic
freedom. As a consequence of scale invariance at large distances, the
theory cannot be in a confining phase and chiral symmetry remains
unbroken. The long-distance dynamics is governed by the IRFP's
critical exponents, which determine the scaling laws in the vicinity
of the fixed point. The range of values $n_{f,\mathrm{lo}}<
n_f<n_{f,\mathrm{up}}$ is known as the ``conformal window''. The
Banks-Zaks theories~\cite{Caswell:1974gg,Banks:1981nn}, where $N_c$
and $n_f$ are arranged such that the critical coupling $g^* \ll 1$,
provide one working example of theories within the conformal window
that can be analysed in perturbation theory.


Theories near the edge of the conformal window (e.g.\ $n_f \lesssim
n_{f,\mathrm{lo}}$) have been recently put forward as candidate
walking theories~\cite{Dietrich:2006cm}. Alternative scenarios start
from a theory inside the conformal window, and deform the theory away
from conformality by perturbing it with operators that are relevant in
the IR regime~\cite{Luty:2004ye}, see also
Refs.~\cite{Dietrich:2005jn,Fukano:2010yv}. In both cases, the
starting point is being able to identify the conformal window, i.e.\
to be able to identify the existence of a fixed point, and to compute
the critical exponents that characterize the relevant directions of
the RG flow. We shall concentrate on this specific problem in this
review.

Informations about the existence of IRFPs in nonsupersymmetric gauge
theories can be obtained by analytical methods like e.g.\ the solution
of the truncated Schwinger-Dyson
equation~\cite{Cohen:1988sq,Appelquist:1996dq,Appelquist:1998rb,Gardi:1998ch},
or the usage of conjectures about the nonperturbative behaviour of
quantum field theories~\cite{Ryttov:2007cx}. This information is
valuable to guide the numerical investigations, but robust results
ultimately need investigations that rely on first principles; besides
the lattice studies discussed below, interesting analytical results
for the anomalous dimensions in a CFT have been obtained in
Ref.~\cite{Rattazzi:2008pe} from first principles. 

\paragraph{Lattice evidence. }

Having set the task to identify the existence of IR fixed points, we
need to spell out clearly what are the observables that can be
evaluated by Monte Carlo methods, and what are the systematic errors
that need to be kept under control in order to draw meaningful
conclusions from the lattice data.

The tools that have been developed for the numerical studies of QCD
can be adapted to this new class of theories, yielding new and
interesting informations. These tools are reviewed in
Sect.~\ref{sec:tools} below.

In order to perform a lattice study, a particular theory has to be
selected and simulated. Analytical results can help in selecting
candidate theories inside the conformal window. However the lack of
robust analytical results means that there is no guarantee of being
inside (or near) the conformal window before simulations are
performed. As a consequence, numerical tools developed for these
studies need to provide:
\begin{itemize}
\item flexibility; so that the same codebase can be used to simulate
  more than just one theory.
\item efficiency; so that simulations are reasonably fast. At this
  stage, it is important to find the right balance between
  optimization and flexibility.
\item a wide range of observables; so that robust conclusions can be
  drawn, based on more than one observation. 
\item detailed numerical benchmarks, so that the algorithmic issues
  can be clearly identified, and distinguished from the physically
  meaningful results. In my opinion these algorithmic studies have not
  yet reached the maturity of their QCD counterparts.
\end{itemize}

\begin{figure}[ht]
  \centering \includegraphics[width=3.5truein]{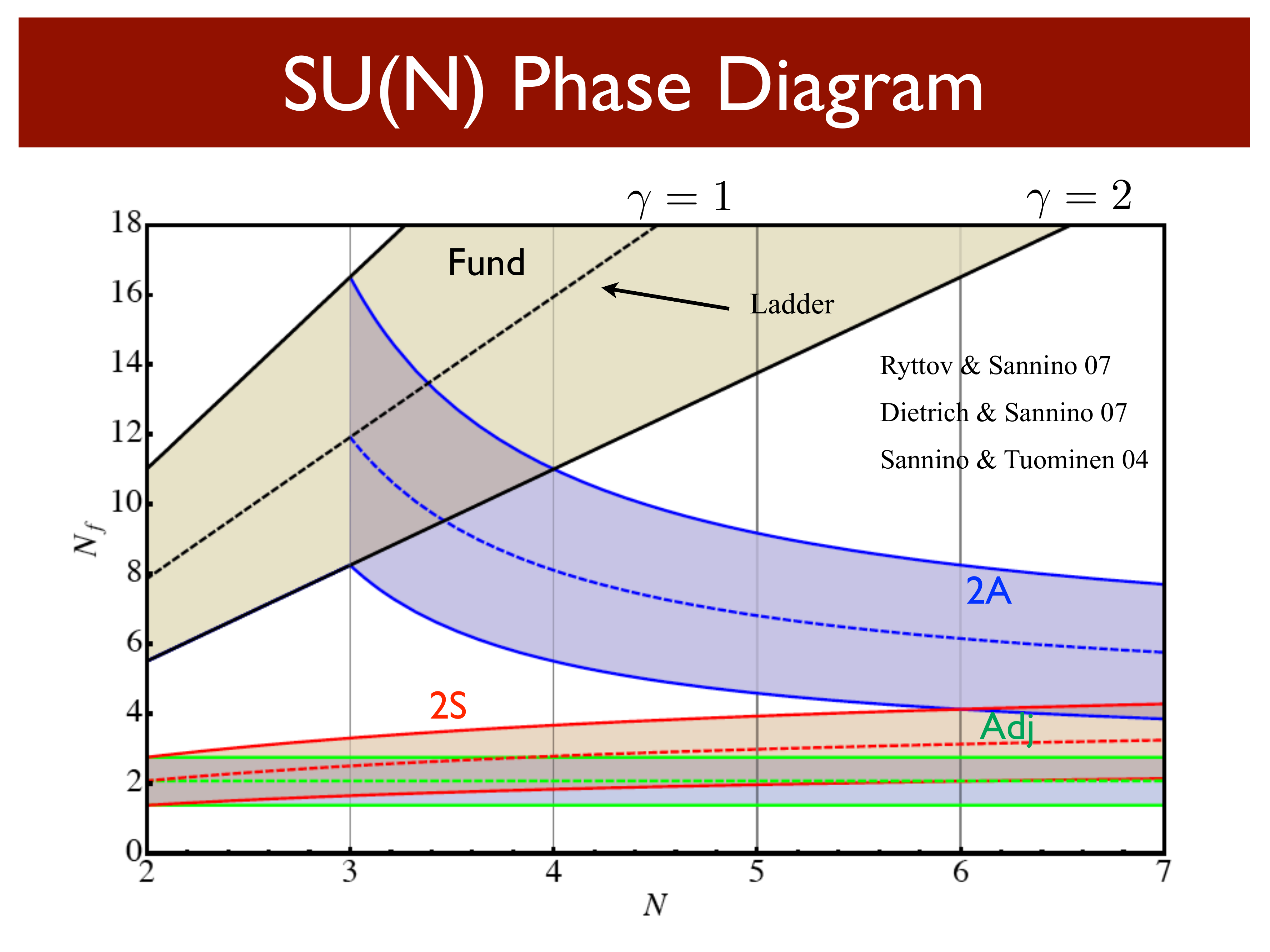}
  \caption{Boundaries of the conformal window for SU(N) gauge theories
    with $n_f$ species of Dirac fermions. The four bands represent
    respectively fermions in the fundamental (Fund), adjoint (A) and
    two-index symmetric and antisymmetric (2S,2A) representations. The
    upper limit of each band corresponds to the number of flavors
    where asymptotic freedom is lost, as obtained from one-loop
    perturbative computations. The lower limit of each band yields the
    number of flavors above which the theories develop an IR fixed
    point. The location of these lower limits relies upon assumptions
    about the nonperturbative dynamics of the theories. Lattice
    simulations can provide first-principle evidence in favour (or
    against) this picture, and compute the critical exponents that
    characterize the fixed points. Figure courtesy of F.~Sannino.}
  \label{fig:CWb}
\end{figure}

A few candidate theories have been studied recently, chosen according
to the conjectured boundaries of the conformal window for SU(N) gauge
theories with fermions in the fundamental, adjoint or two-index
representations. These boundaries have been summarized in
Ref.~\cite{Sannino:2009za}, whence the plot in Fig.~\ref{fig:CWb} is
extracted. As shown in this plot, theories with fermions in higher
dimensional representations of the gauge group are expected to develop
a fixed point for relatively small numbers of flavors. Recent studies
have focused on theories that are close to the lower edge of the
conformal window: SU(3) with $n_f=16,12,10,9,8,6$ flavors in the
fundamental representation, SU(2) with $n_f=6$ flavors in the
fundamental, SU(2) with $n_f=2$ flavors in the adjoint representation,
and SU(3) with $n_f=2$ flavors in the two-index symmetric ({\em
  sextet}) representation. At these early stages of the
nonperturbative studies of the conformal window it is important to try
to identify a paradigm to guide the numerical investigations, rather
than trying to get exhaustive results on one specific theory.

\section{Tools}
\label{sec:tools}

Numerical tools that were originally designed for investigating
lattice QCD have been used in order to identify the existence of
IRFPs. We describe briefly the main ideas, the observables that are
used in the different approaches, and their expected behaviour in the
presence of an IRFP. For each case we try to emphasize the sources of
systematic errors that need to be kept under control in order to draw
robust conclusions from numerical data.

\subsection{Phase structure of the lattice theories. }
\label{sec:phasstr}

Lattice simulations are performed by discretizing the action of a
given theory on a Euclidean space-time lattice. At weak coupling the
RG flow can be computed perturbatively, and the relevant parameters
are easily identified. For an asymptotically-free gauge theory, $g=0$
is an UV fixed point that defines the usual continuum limit of the
lattice theory. The IRFP that we are seeking is a fixed point on the
massless renormalized trajectory that originates from the continuum
limit. As the bare coupling is increased, the lattice theory may have
a complicated phase structure, with new fixed points appearing, which
are not necessarily related to the continuum limit. 

In particular there can be bulk phase transitions in the lattice
theory at strong coupling. These are lattice artefacts and can obscure
the continuum physics that we are interested in. Mapping the phase
structure of the lattice theory is important in order to be able to
identify the genuine features of the continuum limit.

Finite volume effects also play an important role in determining the
long-distance behaviour of a gauge theory. The inverse size of the
lattice $1/L$ is a relevant coupling in the IR, which introduces an
explicit length scale: we can only probe energies above $1/L$ when
performing lattice simulations in a finite box, while the largest
scaling factor that we can probe in a simulation is
$s_\mathrm{max}=L/a$.

There are two potential effects to take into account. On the one hand,
as the box size is decreased, the IR cutoff can potentially change the
behaviour of the gauge theory and drive the theory into different
phases, characterized by the breaking of centre
symmetries~\cite{Narayanan:2007fb,Cossu:2009sq}. It is important to
study the impact of these transitions on the measured observables. On
the other hand, even when the box is large enough and the system is in
the correct phase, $1/L$ determines the size of the scaling
violations. These can be readily included in scaling laws determined
from RG
analyses~\cite{DeGrand:2009mt,DeGrand:2009hu,DelDebbio:2010hu,DelDebbio:2010ze,DelDebbio:2010jy}.

\subsection{Spectral studies}
\label{sec:spec}

Informations about the mass spectrum and the decay constants of a
given gauge theory, can be obtained from the numerical analysis of
two-point correlators, using the standard techniques developed for
QCD. 

Simulations are performed at a finite value of the mass, and in a
finite volume. Both these quantities break conformal invariance at
large distances. The signatures of an IRFP in this framework are the
scaling laws of physical observables as functions of the mass and
volume, as it is often the case when studying critical
phenomena. Using the scaling behaviour of lattice observables to
identify IRFP was proposed in
Refs.~\cite{Luty:2008vs,Sannino:2008pz,DeGrand:2009mt,DelDebbio:2009fd}. Scaling
laws are derived using the RG equations for field
correlators~\cite{DeGrand:2009mt,DelDebbio:2009fd}. A detailed
analysis of the scaling of lattice observables was discussed in
Refs.~\cite{DelDebbio:2010ze,DelDebbio:2010jy}. Scaling laws in the
vicinity of a fixed point were already considered in searches for UV
fixed points in lower-dimensional field theories. For a summary of
results in the case of three-dimensional four-fermi theories, see
e.g. Ref.~\cite{DelDebbio:1997dv} and references therein.

The masses of the states in the technihadron spectrum scale as:
\begin{equation}
  \label{eq:massscal}
  M_H \sim m^{1/y_m}\, ,
\end{equation}
where $y_m=1+\gamma_*$ is the critical exponent associated to the
fermion mass, and $\gamma_*$ is the (scheme-independent) value of the
anomalous dimension {\em at}\ the fixed point.

A generic operator $O$ with appropriate quantum numbers couples to the
states of the theory with strength denoted $G_H$ for a scalar state,
and $F_V$ for a vector one. The scaling of these couplings, which are
related to the decay constant of the state $H$, can also be derived
from the scaling behaviour of two-point correlators. Following the
notation introduced in Ref.~\cite{DelDebbio:2010ze} the scaling
behaviour of the decay constants can be written as:
\begin{equation}
  \label{eq:scalingeq}
  G \sim m^{\eta_G}\, ,~~~~ F\sim m^{\eta_F}\, ,
\end{equation}
with the exponents summarized in Tab.~\ref{tab:exp1}. Note that the
scaling relations hold for any state in the spectrum. In particular
for the pseudoscalar meson states, they yield a modified Banks-Casher
relation. Scaling laws in the vicinity of a fixed point have also been
discussed in the context of the functional RG, see e.g.\
Ref.~\cite{Braun:2009ns,Braun:2010qs}.

\begin{table}[ht]
  \label{tab:exp1}
  \centering
  \begin{tabular}{l|l|l|l|l|r}
    $O$ & {\rm def} & $\matel{0}{ O }{ J^\mathrm{P(C)}(p)
    }$ & $J^\mathrm{P(C)}$ & $\Delta_{O}$  & $\eta_{G[F]}$
    \\[0.1cm] 
    \hline 
    $S$ & $\bar qq$ & $G_{S}$ & $0^{++}$ & $3 - \gamma_*$ & 
    $(2-\gamma_*)/\ym$ \\[0.1cm] 
    $S^a$ & $\bar q \lambda^a q$ & $G_{S^a}$ & $0^{+}$ & $3 - \gamma_*$ &
    $(2-\gamma_*)/\ym$ \\[0.1cm] 
    $P^a$ & $\bar q i\gamma_5 q$ & $G_{P^a}$ & $0^{-}$ & $3 - \gamma_*$ &
    $(2-\gamma_*)/\ym$ \\[0.1cm] 
    $V$ & $\bar q \gamma_\mu q$ & $\epsilon_\mu(p) M_V F_{V} $ &
    $1^{--}$ & $3$ & $1/\ym$ \\[0.1cm]
    $V^a$ & $\bar q \gamma_\mu \lambda^a q$ & $\epsilon_\mu(p) M_V
    F_{V^a} $ & $1^{-}$ & $3 $ & $1/\ym$ \\[0.1cm] 
    $A^a$ & $\bar q \gamma_\mu \gamma_5 \lambda^a q$ & $\epsilon_\mu(p) M_A
    F_{A^a}  $ & $1^{+}$ & $3$ & $1/\ym $   \\[0.1cm] 
    &  & $i p_\mu F_{P^a}$ & $0^{-}$ & $3$ & $1/\ym$  
  \end{tabular} 
  \caption{\small Scaling laws, $G[F]\sim m^{\eta_{G[F]}}$ for decay constants.
    The symbol $\ym \equiv 1 + \gamma_*$ denotes the
    scaling dimension of the mass 
    and $\Delta_{O} = d_{O} + \gamma_{O}$ is the
    scaling dimension of the operator $O$. 
    The symbol $a$ denotes the adjoint flavour index, and $\lambda^a$
    are the generators normalized as $\mathrm{tr}[\lambda^a \lambda^b]
    = 2 \delta^{ab}$. No such simple expression exists for
    the axial singlet current because
    of the chiral anomaly~\cite{DelDebbio:2010ze}.}  
\end{table}

It is important to bear in mind that scaling laws are asymptotic
formulae that are obtained by linearizing the RG equations in a
neighbourhood of the fixed point at $m=0$ and $1/L=0$. As the system
moves towards larger values of $m$ and $1/L$, the corrections to
scaling become important and eventually obscure the power-law scaling
dictated by the IRFP. The critical exponents can be extracted from the
power-law scaling of the spectrum only from simulations at small mass.

The distinctive feature in the spectrum of an IR-conformal theory is
the lack of spontaneous chiral symmetry breaking. As a consequence,
there are no Goldstone bosons in the theory. As the fermion mass is
sent to zero {\em all}\ states in the spectrum become massless, see
e.g. Ref.~\cite{Miransky:1998dh} for a realization of this
scenario. This is at odds with the chirally-broken scenario, where
there is a parametric separation between the pseudoscalar Goldstone
bosons and the rest of the massive spectrum. Any effective theory that
describes the low-energy dynamics of an IR-conformal system must take
all the light degrees of freedom into account.

As simulations move to smaller fermion masses, the physical size of
the lattice must be increased in order to avoid finite-size
effects. In particular a chirally broken theory at small mass and
fixed physical volume can enter the so-called $\delta$-regime, where
the spectrum is determined by the rotator states of the chiral
condensate~\cite{Leutwyler:1987ak}. This regime corresponds to $L_t\gg
L_s$, $F L_s>1$, and $M L_s \ll 1$; the spectrum of the lowest states
no longer scales as expected for the Goldstone bosons of a chirally
broken theory. In this case deviations from the GMOR scaling are not a
signal of restoration of chiral symmetry. Similarly if $L_t\sim L_s$
the system will be driven in the $\epsilon$ regime, which can be seen
as the high-temperature limit of the $\delta$-regime. Great care must
be exercised in the interpretation of lattice data, as these finite
size effects can easily be mistaken for signals of conformal
behaviour. This has been clearly pointed out in
Ref.~\cite{Fodor:2009wk}. The $\delta$ regime can be distinguished
from a conformal one if the power-law scaling corresponding to the
latter is observed.

Last but not least, it is important to verify that the simulated
masses do not correspond to the ``heavy quark'' limit of a chirally
broken theory. This can be achieved by comparing the mesonic and
gluonic sectors of the spectrum as discussed in the next section, where
we report the numerical results. 

\subsection{SF studies}
\label{sec:SF}

The nonperturbative running of the gauge coupling and fermion mass can
be studied using the Schr\"odinger functional
scheme~\cite{Luscher:1991wu,Luscher:1992an}. The running coupling
$\overline{g}^2$ at the scale $1/L$ is defined on a hypercubic lattice
of size $L$, with boundary conditions chosen to impose a background
chromoelectric field, which depends on a parameter $\eta$. The
renormalized coupling is defined as a measure of the response of the
system to changes in the background chromoelectric field:
\begin{equation}
  \label{eq:SFcoupling}
  \overline{g}^2=k \left< \frac{\partial S}{\partial \eta}
  \right>^{-1}\, ,
\end{equation}
where $S$ is the action of the Schr\"odinger functional, and the
constant $k$ is chosen such that $\overline{g}^2=g_0^2$ to leading
order in perturbation theory. Eq.~(\ref{eq:SFcoupling}) defines a
nonperturbative coupling, which depends on only one scale, the size of
the system $L$, and can be evaluated numerically.

To measure the running of the quark mass, we calculate the
pseudoscalar density renormalisation constant $Z_P$. Following
Ref.~\cite{Capitani:1998mq}, $Z_P$ is defined by:
\begin{equation}
  \label{eq:ZPdef}
  Z_P(L)=\sqrt{3 f_1}/f_P(L/2)\, ,
\end{equation}
where $f_1$ and $f_P$ are the correlation functions involving the
boundary fermion fields $\zeta$ and $\overline{\zeta}$:
\begin{eqnarray}
  \label{eq:f1def}
  f_1&=&-1/12L^6 \int d^3u\, d^3v\, d^3y\, d^3z\,
  \langle
  \overline{\zeta}^\prime(u)\gamma_5\tau^a{\zeta}^\prime(v)
  \overline{\zeta}(y)\gamma_5\tau^a\zeta(z) 
  \rangle\, , \\
  \label{eq:fPdef}
  f_P(x_0)&=&-1/12 \int d^3y\, d^3z\,  \langle
  \overline{\psi}(x_0)\gamma_5\tau^a\psi(x_0)\overline{\zeta}(y)\gamma_5\tau^a\zeta(z)
  \rangle\, .
\end{eqnarray}
For the details of the Schr\"odinger functional setup we refer the
reader to the original publications.

The running of the coupling as the scale is varied by a factor $s$ is
encoded in the step scaling function $\sigma(u,s)$ as
\begin{eqnarray}
  \label{eq:Sig}
  \Sigma(u,s,a/L)&=&\left. \overline{g}^2(g_0,sL/a)
  \right|_{\overline{g}^2(g_0,L/a)=u}\, , \\
  \label{eq:sigma}
  \sigma(u,s) &=& \lim_{a/L\to 0} \Sigma(u,s,a/L)\, ,
\end{eqnarray}
as described in Ref.~\cite{Luscher:1992an}. The function $\sigma(u,s)$
is the continuum extrapolation of $\Sigma(u,s,a/L)$ which is obtained
from numerical simulations at various $a/L$ values and fixed $u$.  The
step scaling function encodes the same information as the $\beta$
function. The relation between the two functions for a generic
rescaling of lengths by a factor $s$ is given by:
\begin{equation}
  \label{eq:stepbeta}
  -2 \log s = \int_{u}^{\sigma(u,s)} \frac{dx}{\sqrt{x}
    \beta(\sqrt{x})}\, .
\end{equation}
The step scaling function can be computed at a given order in
perturbation theory by using the analytic expression for the
perturbative $\beta$ function, and solving Eq.~(\ref{eq:stepbeta}) for
$\sigma(u,s)$.  It can be seen from the definition of $\sigma(u,s)$ in
Eq.~(\ref{eq:sigma}) that an IRFP corresponds to $\sigma(u,s)=u$.

The lattice step scaling function for the mass is defined as:
\begin{equation}
  \label{eq:sigmaPdef}
  \Sigma_P(u,s,a/L)=\left
    . {\frac{Z_P(g_0,sL/a)}{Z_P(g_0,L/a)}}
  \right|_{\overline{g}^2(L)=u}\, ; 
\end{equation}
the mass step scaling function in the continuum limit,
$\sigma_P(u,s)$, is given by:
\begin{equation}
  \label{eq:sigma_p}
  \sigma_P(u,s) = \lim_{a\to 0}\Sigma_P(u,s,a/L)\, .
\end{equation}
The mass step scaling function is related to the mass anomalous
dimension (see e.g. Ref.~\cite{DellaMorte:2005kg}):
\begin{equation}
  \label{eq:sigPgamma}
  \sigma_P(u) = \left(\frac{u}{\sigma(u)}\right)^{(d_0/(2\beta_0))}
  \exp\left[\int_{\sqrt{u}}^{\sqrt{\sigma(u)}} dx 
    \left(\frac{\gamma(x)}{\beta(x)}-\frac{d_0}{\beta_0 x}\right)\right]\, .
\end{equation}

In the vicinity of an IRFP the relation between $\sigma_P$ and
$\gamma$ simplifies:
\begin{equation}
  \label{eq:gammafix}
  \int_{\overline{m}(\mu)}^{\overline{m}(\mu/s)} \frac{dm}{m} = 
  -\gamma_* \int_{\mu}^{\mu/s} \frac{dq}{q}\, ,
\end{equation}
and hence:
\begin{equation}
  \label{eq:sigmaPfix}
  \log\left|\sigma_P(s,u)\right| = -\gamma_* \log s\, . 
\end{equation}
We can therefore define an estimator
\begin{equation}
\label{eq:tau}
\hat\gamma(u) =
-\frac{\log\left|\sigma_P(u,s)\right|}{\log\left|s\right|}\, ,
\end{equation}
which yields the value of the anomalous dimension at the fixed
point. Away from the fixed point $\hat\gamma$ will deviate from the
anomalous dimension, with the discrepancy becoming larger as the
anomalous dimension develops a sizeable dependence on the energy
scale. 

The step scaling functions need to be extrapolated to the continuum
limit in order to disentangle the running of the couplings from the
lattice artefacts that affect the measured observables. This
extrapolation is very delicate for $\sigma(u,s)$.  In the vicinity of
a fixed point the running is by definition very slow. Therefore a very
high accuracy in the extrapolation is needed in order to resolve the
physically meaningful signal. The systematic uncertainties make it
difficult to locate the value of the critical coupling with sufficient
accuracy. In the next section, we shall discuss in more detail the
propagation of errors and their impact on the physically interesting
results.

\subsection{Potential schemes}
\label{sec:potscheme}

Finally the running of the coupling can be studied by defining a
nonperturbative coupling from the potential between static charges
computed numerically~\cite{Bilgici:2009kh}.

The coupling is defined as:
\begin{equation}
  \label{eq:potcoup}
    g_w^2(L_0,R/L_0,a/L_0) = \frac{1}{k}(R/a)^2 \chi(R/a,L_0/a)\, ,
\end{equation}
where $\chi$ is a Creutz ratio, $L_0$ is the size of the lattice, and
$R$ is the size of the Wilson loops used to construct the Creutz
ratios.  The size of the lattice $L_0$ sets the scale at which the
coupling is defined, while the ratio $R/L_0$ defines the
renormalization scheme. 

A step scaling function can be defined in close analogy to the one
defined for the Schr\"odinger functional:
\begin{eqnarray}
  &&\Sigma_w(u,R/L_0,a/L_0) = g_w^2(bL_0,R/L_0,a/L_0)\, ,\\
  && u=g_w^2(L_0,R/L_0,a/L_0) \, ,\\
  && \nonumber \\
  \label{eq:last}
  && \sigma_w(u) = \lim_{a\to 0} \Sigma_w(u,R/L_0,a/L_0) \, .
\end{eqnarray}

Once again the extrapolation to the continuum limit in
Eq.~(\ref{eq:last}) is necessary in order to avoid contaminations from
lattice artefacts.

Ref.~\cite{Fodor:2009rb} presents a variation on the same theme. The
potential between static charges is measured by lattice simulations,
and is compared with the results obtained from integrating the running
coupling: 
\begin{equation}
  \label{eq:Vint}
  V(R) - V(R_0) = C_s(R) \int_{R_0}^R dR^\prime\,
  \frac{\alpha(R^\prime)}{R^{\prime 2}}\, ,
\end{equation}
where the perturbative expansion is used as an input for $\alpha(R)$
in the integral. This method allows one to compare the nonperturbative
running with the perturbative expectation.

\subsection{MCRG}
\label{sec:mcrg}

Monte Carlo Renormalization Group (MCRG) methods were developed in
order to study the coupling flow in both spin and gauge models. In
particular, the 2-lattice matching has proved to be useful in pure
Yang-Mills
theories~\cite{Hasenfratz:1984hx,Hasenfratz:1984bx,Bowler:1984hv}.

The basic idea is to follow the RG flow of the bare couplings under
blocking transformations that integrate out the UV degrees of
freedom. With each blocking step, changing the scale by a factor $s$,
the flow drives the couplings towards a lower-dimensional manifold,
whose dimension is given by the number of relevant couplings. The
distance from this manifold goes as a power of $s$.

\subsubsection{Two-lattice matching procedure}

\begin{figure}[ht]
  \centering
  \includegraphics[width=0.6\textwidth]{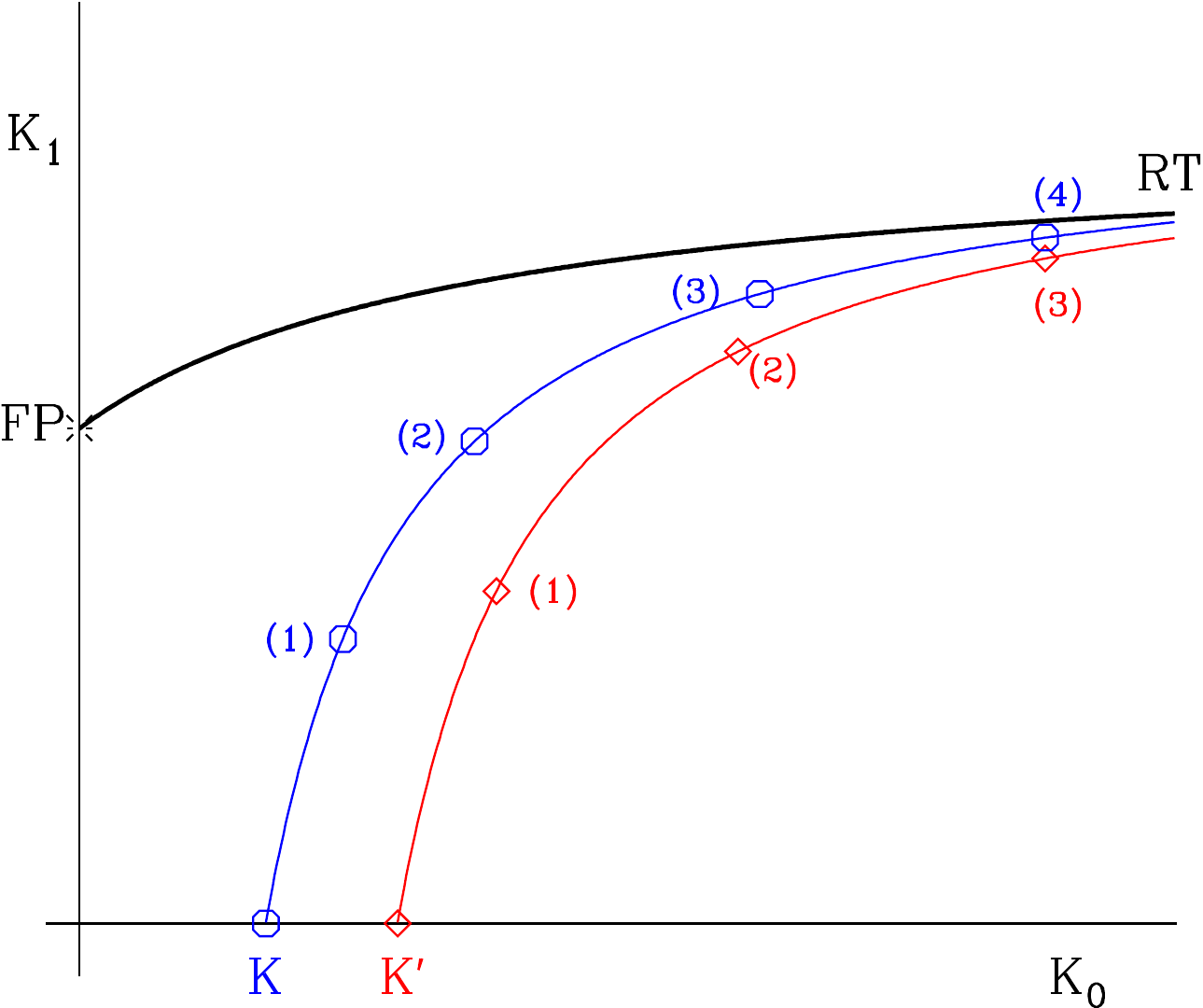}
  \caption{RG flow in bare parameter space. The points correspond to
    successive blocking steps. The figure is taken from
    Ref.~\cite{Hasenfratz:2009ea}.  }
  \label{fig:MCRGflow}
\end{figure}

Let us consider first for simplicity a theory that has only one
relevant parameter flowing out from an UV fixed point; this is the
common situation in pure gauge theories. In this case the RG
trajectories converge towards a one-dimensional renormalized
trajectory (RT). Starting from a value $g$ of the bare coupling at the
cutoff scale, after $n$ steps the Wilsonian action is described by
some point in parameter space, for a sufficiently large $n$ this point
is close to the RT. This is represented by the circles and the blue
curve in Fig.~\ref{fig:MCRGflow}, for $n=4$ steps of blocking starting
from the value $g=K$. This point is {\em matched} by the point
$g^\prime=K^\prime$ such that the RG flow from $K^\prime$ ends up at
the same point as the previous flow after only $n-1$ steps. This flow
is represented by the diamonds on the red trajectory for $n=3$. Since
the two trajectories end at the same point, the lattice correlation
length at the endpoint of the two flows must be the same for both
theories: $\hat \xi(g)/s^n=\hat \xi(g^\prime)/s^{n-1}$, and thus:
\begin{equation}
  \label{eq:RGfow}
  a(g^\prime) = s a(g)\, .
\end{equation}

To identify such a pair of couplings, we need to show that after $n$
and $(n-1)$ RG steps respectively their actions are
identical. Explicitly calculating the actions would be complicated,
but instead the gauge configurations themselves can be blocked;
showing that the expectation values of all observables on these gauge
configurations agree is equivalent to directly comparing the actions
that generated them.

This procedure identifies, for each $n$, a pair of bare gauge couplings
$(g,g^\prime)$, or equivalently $(\beta,\beta')$ where $\beta=2N/g^2$,
with lattice correlation lengths that differ by a factor $s$,
$\hat{\xi}' = \hat{\xi}/s$. In the limit $n \rightarrow \infty$, it is
customary to define the step scaling function for the bare coupling:
$\Delta \beta = \beta - \beta' \equiv s_{b}(\beta;s)$. This is the
analogue of the Schr\"odinger Functional step scaling function for the
renormalised coupling, $\sigma(u,s)$, and in the UV limit where
$\overline{g}^2 \rightarrow g_0^2 = 2N/\beta$, there is a simple
relation between the two:
\begin{equation}
\frac{s_{b}(\beta;s)}{\beta} = \frac{\sigma(u,s)}{u} - 1\, .
\end{equation}
Clearly an IRFP is found when $s_b(\beta^*;s)=0$, while $s_b$ is
expected to remain positive for a QCD-like theory.

There is a degree of arbitrariness in choosing the blocking
transformation, see e.g.\ Ref.~\cite{Hasenfratz:2009ea}. Recent
studies have used:
\begin{equation}
  V_{n,\mu} =
  \mathrm{Proj}\left[(1-\alpha)U_{n,\mu}U_{n+\mu,\mu}+
    \frac{\alpha}{6}\sum_{\nu\neq\mu}U_{n,\nu}U_{n+\nu,\mu}
    U_{n+\mu+\nu,\mu}U_{n+2\mu,\nu}^{\dagger}\right]\, ,
\end{equation}
where $\alpha$ is a free parameter, which can be varied to optimise
the transformation. Another possible choice for blocking is to perform
a so-called
HYP-smearing~\cite{Hasenfratz:2010fi,Hasenfratz:2001hp}. Changing the
blocking transformation changes the location of the fixed point, and
the rate of convergence towards the RT.  Ideally it should be chosen
such that: (i) All observables yield the same $(g,g^\prime)$ pairs for
a given number of blocking steps $n$. Deviations are a
measure of the systematic error from not being at exactly the same
point along the RT; (ii) consecutive blocking steps predict the same
matching coupling, i.e. for a given $g$, the matching coupling
$g^\prime$ should be the same for all $n$. Deviations show that the
irrelevant couplings still have sizeable effects. MCRG yields robust
information only if these systematic errors are under control. 

\section{Results 2010}
\label{sec:res}

We shall summarize here the latest results at the time of the Lattice
Conference. Previous studies have been summarized in the plenary talks
at the Lattice Conferences in 2008 and
2009~\cite{Fleming:2008gy,Pallante:2009hu}.

\subsection{SU(3) with fundamental fermions}
\label{sec:su3f}

Starting from the upper end of the conformal window $n_f=16$, several
theories have been studied with decreasing numbers of fermions. 

\begin{figure}[ht]
  \centering
  \includegraphics[width=0.4\textwidth]{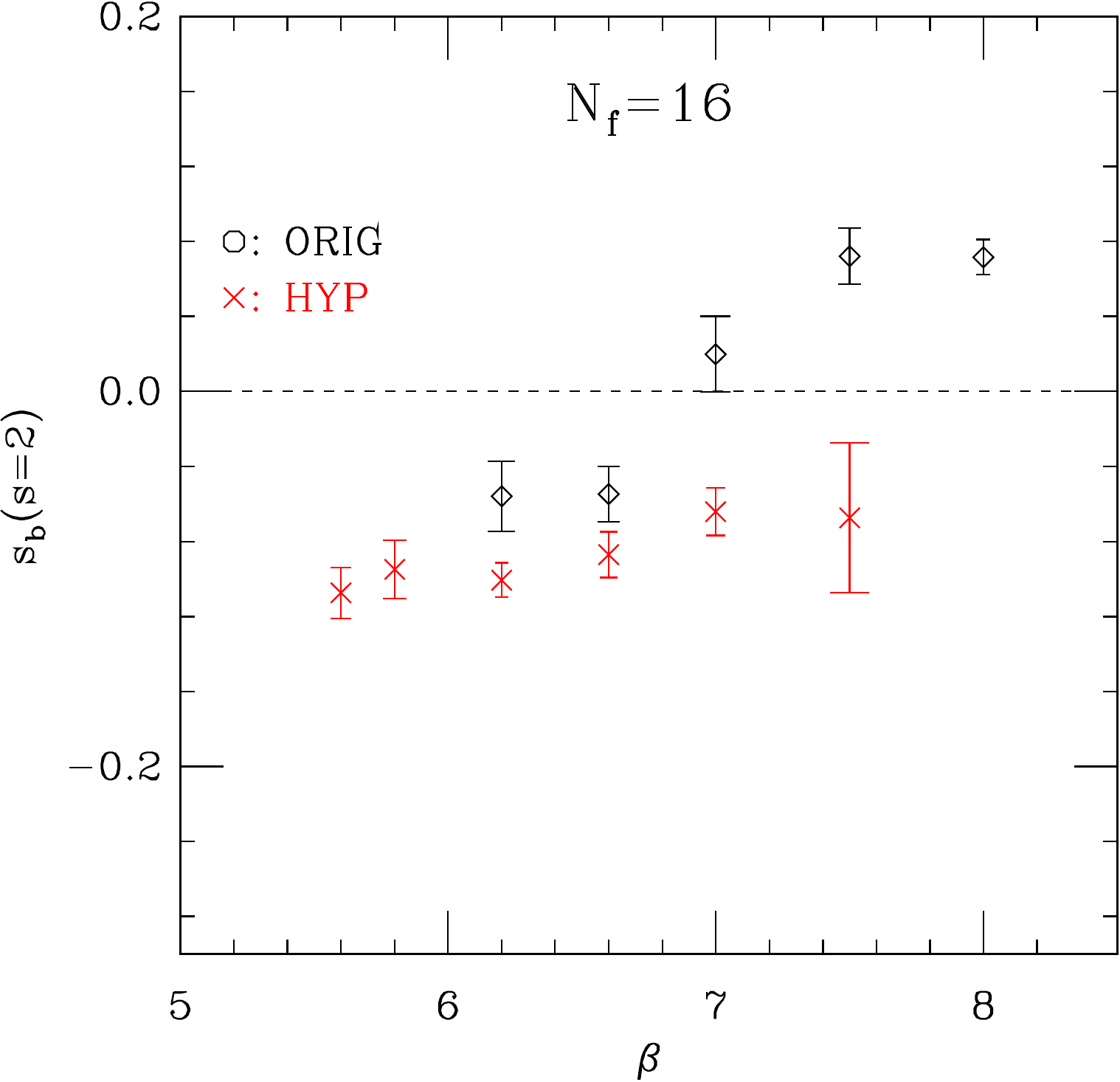}
  \includegraphics[width=0.39\textwidth]{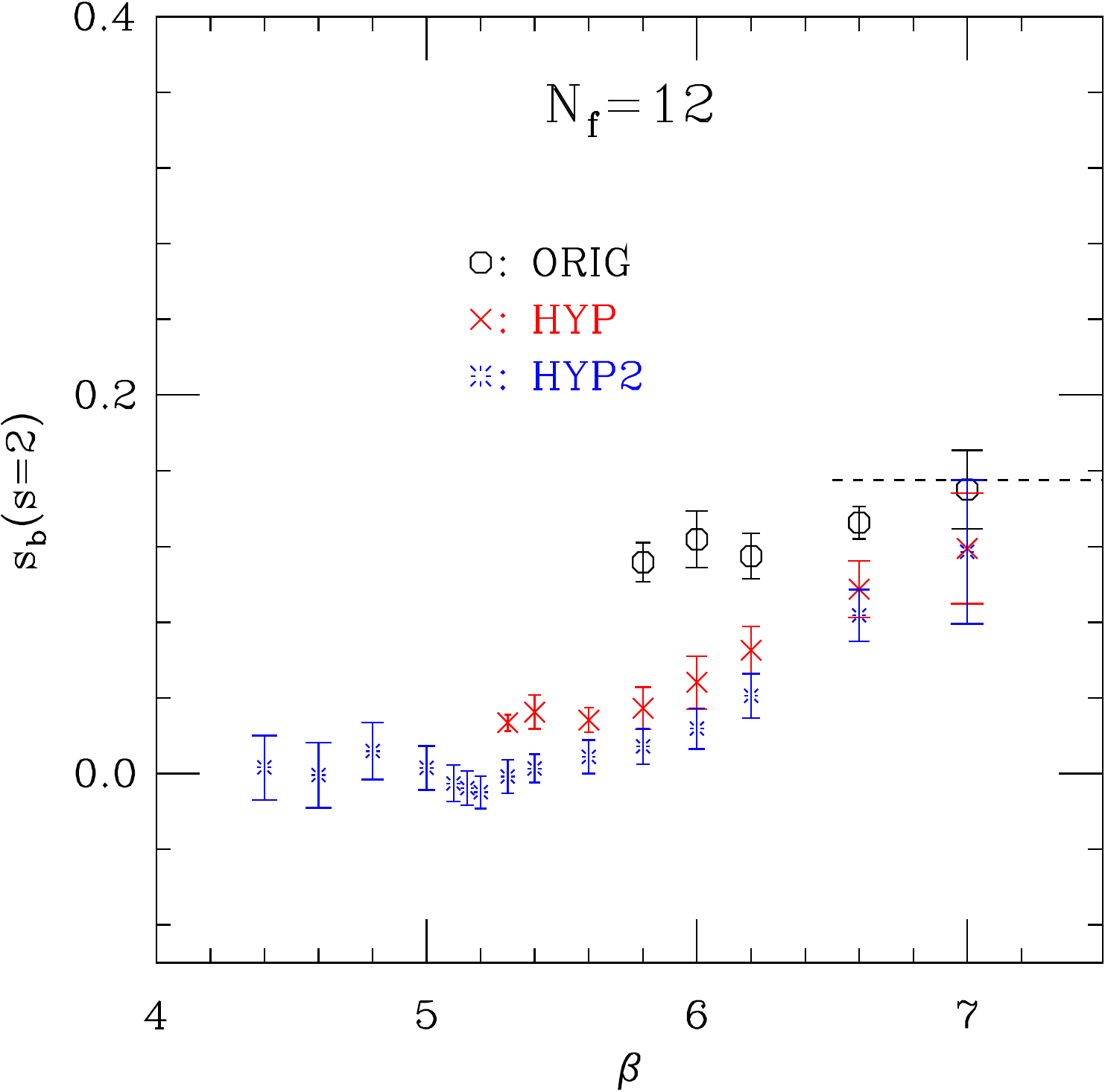}
  \caption{Bare step scaling function for $n_f=16$ (left) and $n_f=12$
    (right) using the {\tt ORIG} ($\diamond$), and the {\tt HYP}
    ($\times$) blocking procedures; details about the blocking
    procedure can be found in Ref.~\cite{Hasenfratz:2010fi}.  }
  \label{fig:su3mcrg}
\end{figure}

At $n_f=16$ the two-loop beta function predicts an IRFP at weak
coupling, $g_0^2\approx 0.5$. MCRG studies of this
theory~\cite{Hasenfratz:2010fi} have found a negative step scaling
function $s_b$, in agreement with the hypothesis that the theory is
indeed inside the conformal window. The bare step scaling function is
shown to the left of Fig.~\ref{fig:su3mcrg}. Different types of
blocking yield different locations of the zero of the step scaling
function, as expected since the position of the fixed point depends on
the renormalization (or blocking) scheme.

At $n_f=12$ the situation is less clear. MCRG studies suggest a slow
running of the coupling, but do not provide a clear cut answer about
the existence of a fixed point. The bare step scaling is shown on the
right of Fig.~\ref{fig:MCRGflow}. The main limitation is the fact that
these simulations require to run the code at exceedingly large values
of the bare coupling~\cite{Hasenfratz:2010fi}. Larger lattices could
help to improve these results by allowing more blocking steps.

Studies of the running of the coupling defined through potential
schemes do not show any sign of a conformal fixed point in this
case. We refer the reader to the talks of Itou and Holland in these
Proceedings~\cite{Itou:lat10,Holland:lat10} for more details on these
studies, which disagree with the results presented in
Refs.~\cite{Appelquist:2007hu,Appelquist:2009ty}.

\begin{figure}[ht]
  \centering
  \includegraphics[width=0.7\textwidth]{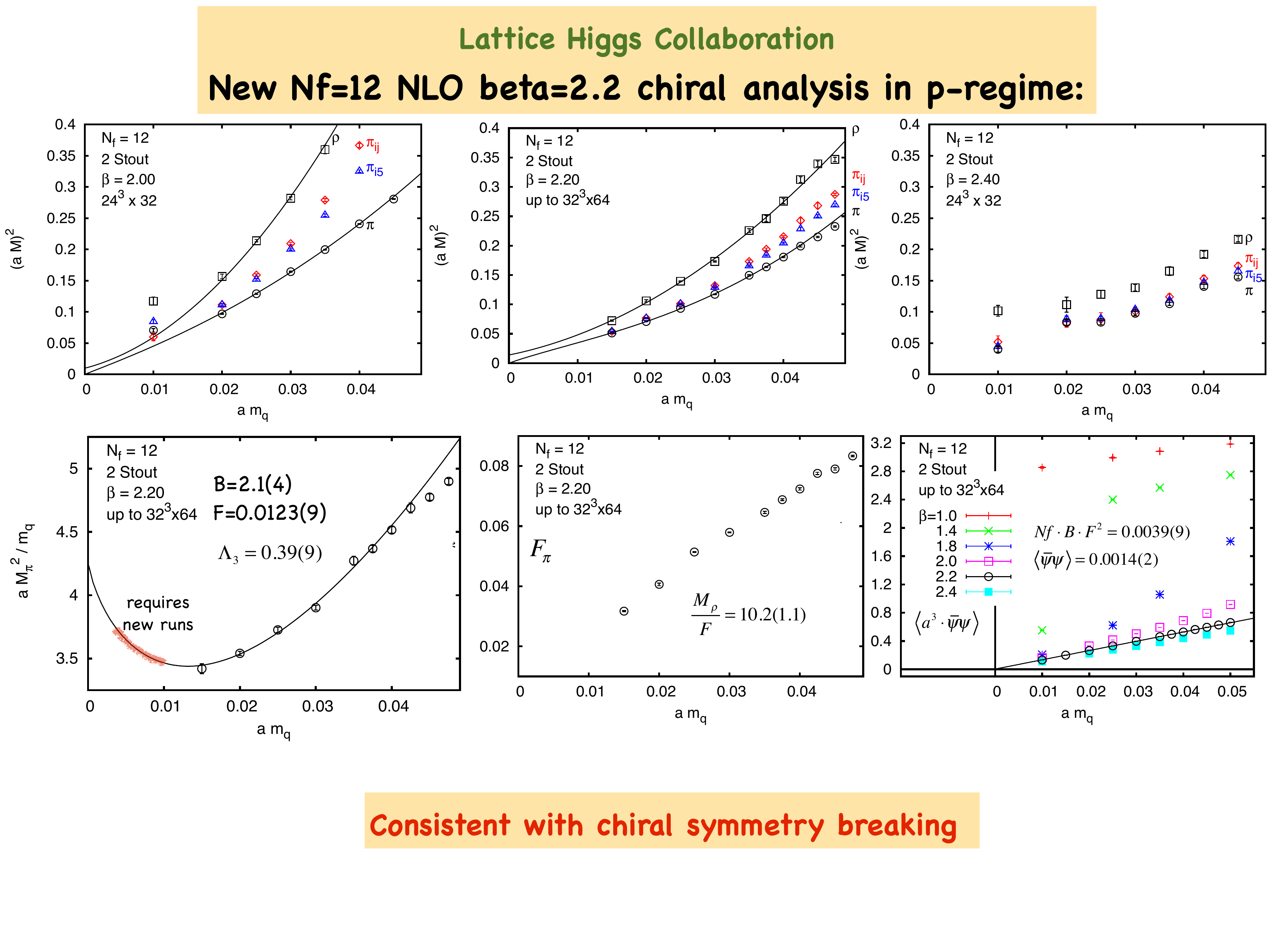}
  \caption{Spectrum studies for SU(3) with $n_f=12$. The upper plots
    show the pseudoscalar masses. The lower plots display, from left
    to right, the GMOR relation, the pion decay constant, and the
    chiral condensate. Lines correspond to fits to chiral perturbation
    theory~\cite{Kuti:lat10}.}
  \label{fig:su3spec}
\end{figure}

Spectral studies for this theory have not reached consensus yet. Some
are consistent with the spectrum of a confining
theory~\cite{Kuti:lat10,Jin:lat10,Fodor:2009wk}. A summary of these
results is reported in Fig.~\ref{fig:su3spec}. However other studies
fit well the hypothesis that the theory is in a chirally symmetric
regime~\cite{Lombardo:lat10,Pallante:lat10,Deuzeman:2009mh,Deuzeman:2009ze}. As
discussed in the previous section, there are a number of systematic
errors that are likely to obscure the physically meaningful results,
and more extensive simulations will be required to settle this issue.

As $n_f$ is further decreased, results for $n_f=10,9,8$
indicate these theories are already below the edge of the conformal
window~\cite{Hasenfratz:2010fi,Fodor:2009wk}. 

Recent results, which appeared after the Lattice conference, suggest
instead that the theory with $n_f=10$ lies inside the conformal
window~\cite{Hayakawa:2010yn}.

The LSD collaboration has been studying the $n_f=6$ theory. This
theory is expected to be in the confining regime, sufficiently close
to the edge of the conformal window to display a walking behaviour. An
enhancement of the chiral condensate by a factor of 2 at the cutoff
scale, and a possible trend towards parity doubling for the mesonic
states have been identified~\cite{Fleming:lat10}. The phenomenological
implications of these findings need to be studied in more detail.

\subsection{SU(2) with adjoint fermions}
\label{sec:su2adj}

The SU(2) gauge theory with $n_f=2$ flavors in the adjoint
representation has been investigated by several groups using different
methods. In this case all simulations seem to indicate that the theory
is inside the conformal window.

\begin{figure}[ht]
  \centering
  \includegraphics[width=2.4truein,height=2truein]{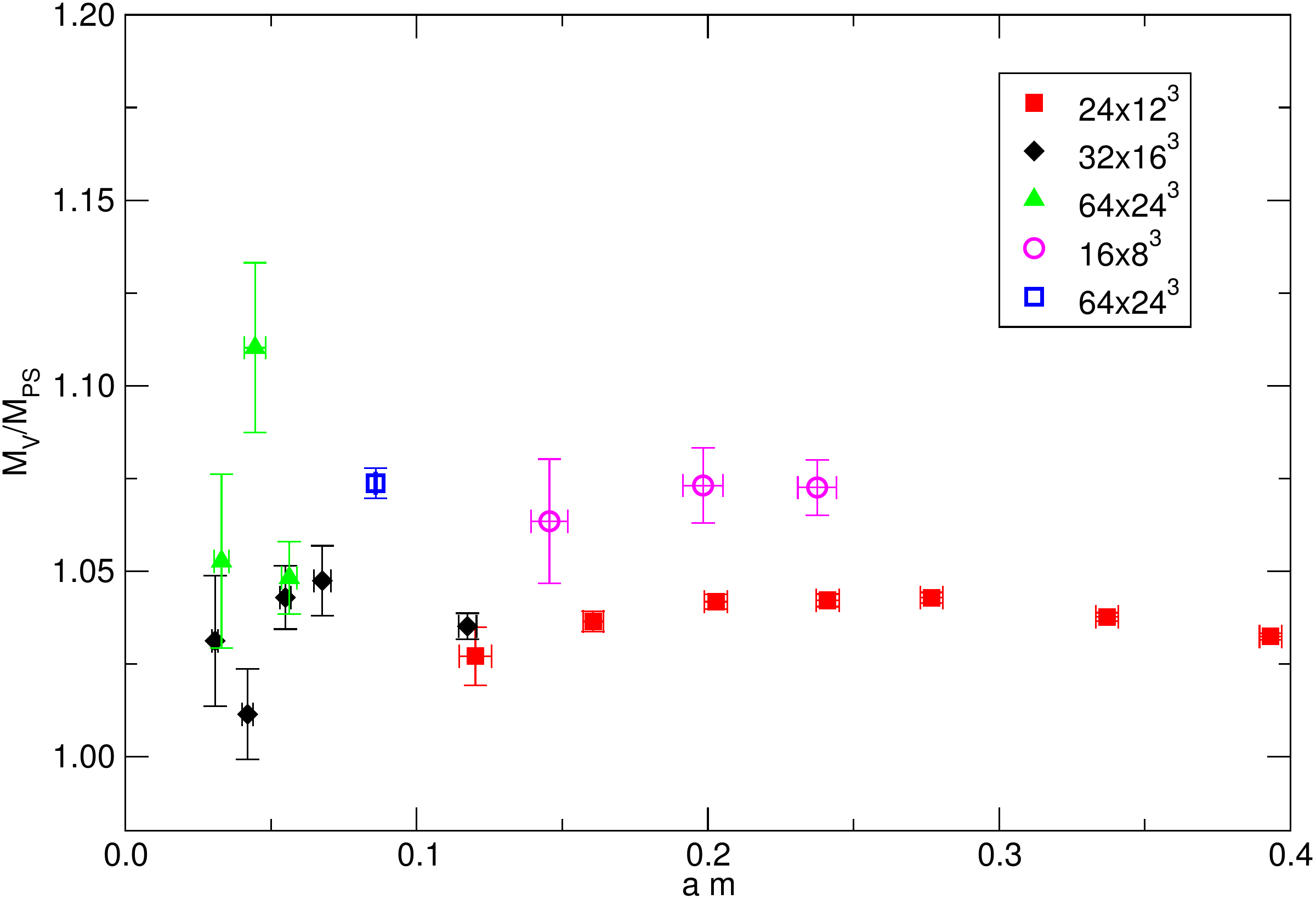}
  \includegraphics[width=2.35truein,height=2truein]{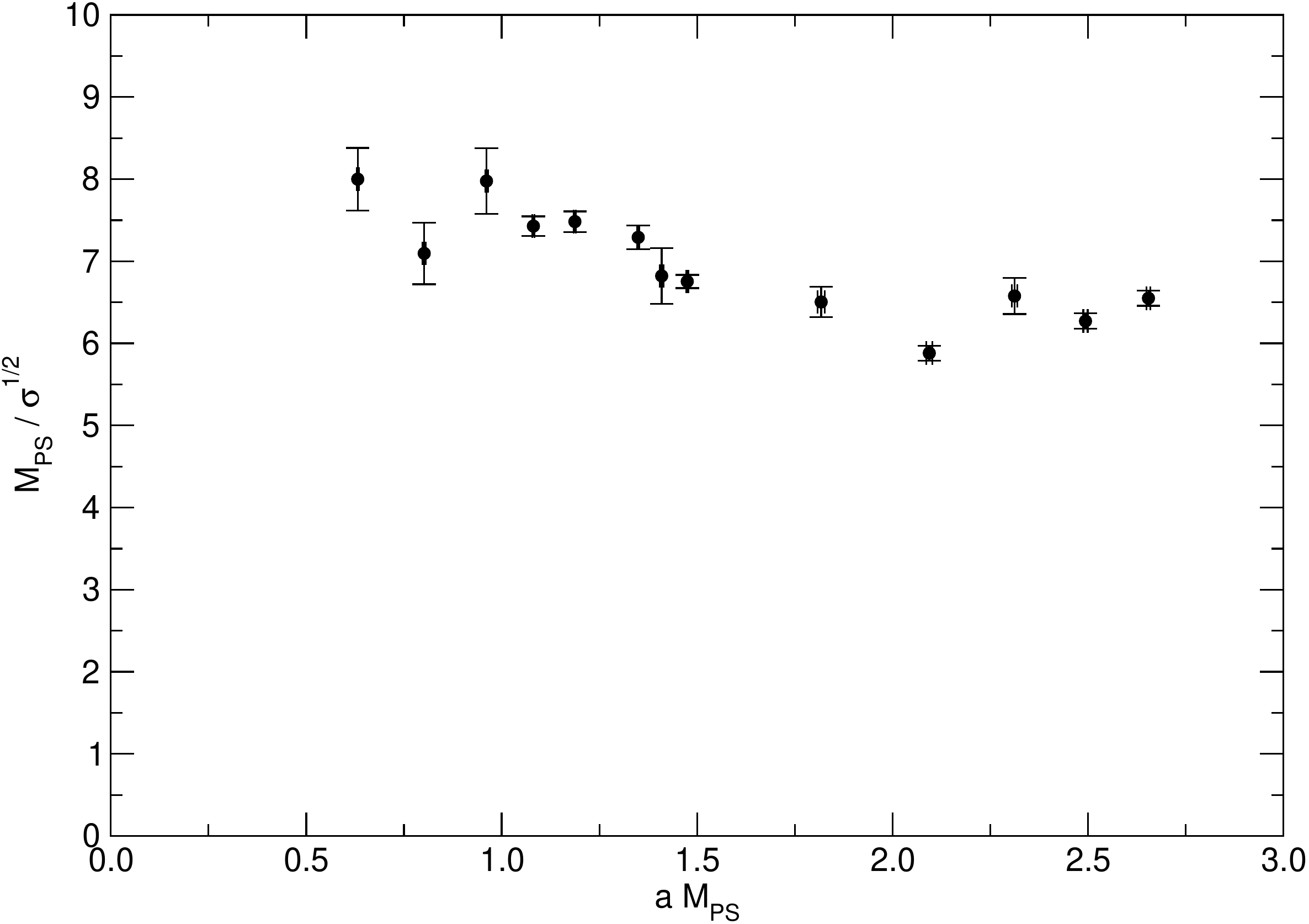}
  \caption{Vector meson to pseudoscalar meson mass ratio (left);
    notice that the two states remain degenerate when the mass is
    decreased. Ratio of the pseudoscalar mass to the square root of
    the string tension (right). The pseusoscalar mass and the string
    tension vanish at the same rate, yielding a finite ratio.}
  \label{fig:su2spec}
\end{figure}

Recent studies are reported in
Refs.~\cite{Catterall:2007yx,DelDebbio:2008wb,DelDebbio:2008zf,Catterall:2008qk,DelDebbio:2008tv,Hietanen:2008mr,Hietanen:2008vc,DelDebbio:2009fd,Pica:2009hc}. Simulations
so far have been performed with non-improved Wilson fermions; the
phase diagram for the lattice theory has been mapped carefully in
Refs.~\cite{Hietanen:2008mr,Catterall:2008qk}, where a bulk phase
transition was found and the region connected to continuum physics has
been identified. Simulations have been performed trying to reach the
small mass regime while preserving the hierarchy of scales required to
control the systematic errors:
\begin{equation}
  \label{eq:scalehier}
  \left(\frac{L}{a}\right) \ll a m_\mathrm{PS} \ll 
  \left(\frac{r_0}{a}\right)^{-1} \ll 1\, ,
\end{equation}
where $L$ is the lattice size, $m_\mathrm{PS}$ is the lightest mass in
the mesonic spectrum, and $r_0$ is the Sommer radius. Satisfying the
above inequalities at small fermion masses becomes very rapidly a
CPU-intensive task. New results for the spectrum were presented at
this
Conference~\cite{Pica:lat10,Patella:lat10,Kerrane:lat10}. Fig.~\ref{fig:su2spec}
summarizes the most striking features observed in the spectrum, namely
the near-degeneracy of the pseudoscalar and the vector meson mass,
which persists at the smallest masses explored so far, and the large
ratio of the pseudoscalar mass to the string tension. Both behaviours
are at odds with the expected behaviour in a theory where chiral
symmetry is spontaneously broken. In the latter case, the pseudoscalar
mass goes to zero, while the other two quantities remain finite, thus
yielding respectively a divergent and a vanishing ratio for the
quantities in Fig.~\ref{fig:su2spec}. Another interesting aspect of
the spectrum study presented in Ref.~\cite{DelDebbio:2010hx} is the
hierarchy between the glueball and the mesonic states, with the former
being lighter than the latter. The mass of the glueballs scales with
the fermion mass $m$, indicating that the system is not in the heavy
mass regime. This result suggests that the light glueball states need
to be included in any effective lagrangian description of TC
low-energy dynamics. Note that these results have been obtained at a
single value of the lattice spacing and therefore the size of lattice
artefacts cannot be estimated properly.

\begin{figure}[ht]
  \centering
  \includegraphics[width=0.6\textwidth]{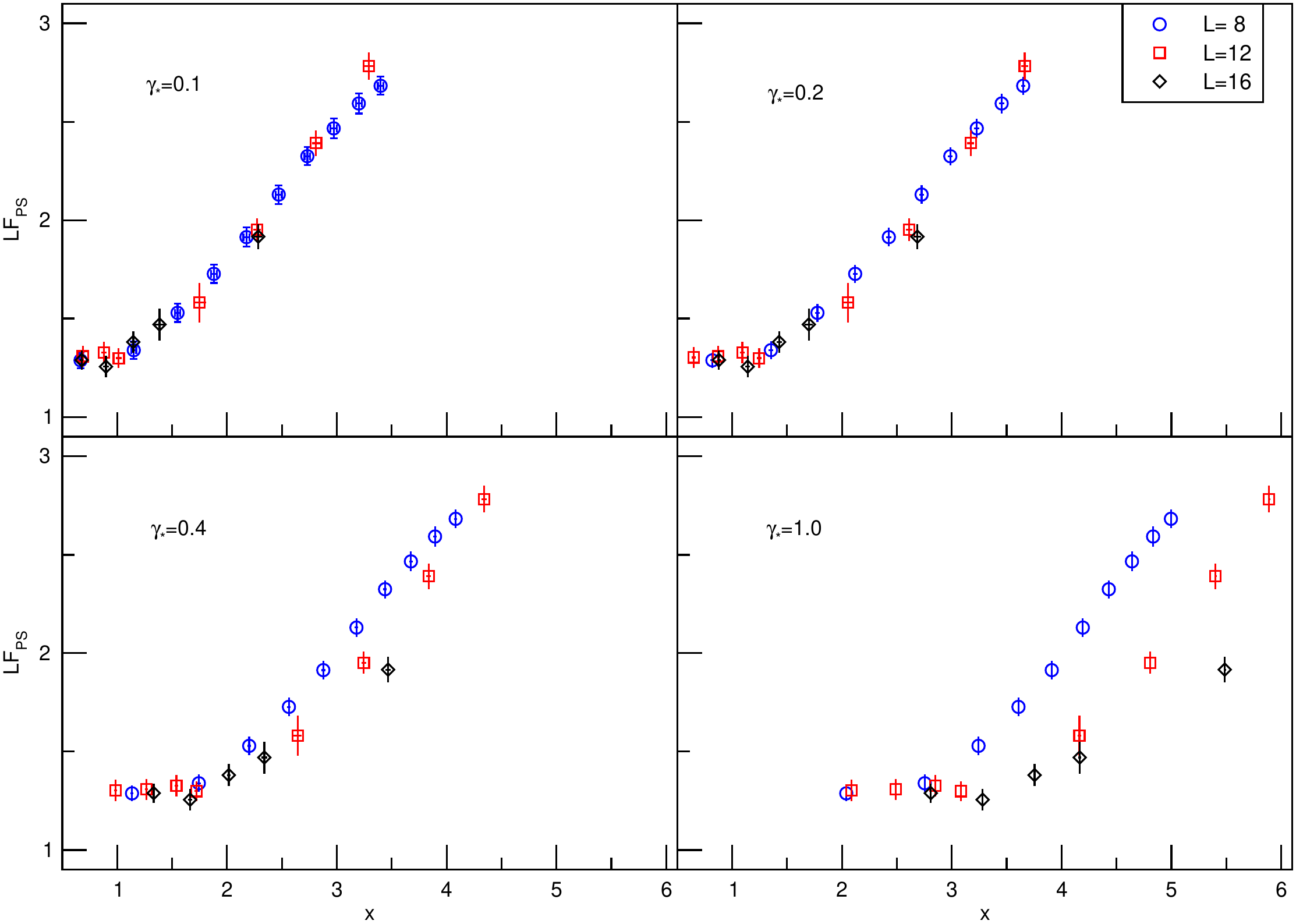}
  \caption{Finite size scaling curves for the pseudoscalar decay
    constant. The four plots, clockwise from the top left corner,
    correspond to scaling exponents $\gamma_*=0.1,0.2,0.5,0.4$
    respectively. }
  \label{fig:su2fss}
\end{figure}

Finite size scaling can be tested by rescaling data obtained on
lattices of varying size. Assuming the existence of an IRFP, data from
different lattices should fall on a universal curve, e.g.\ for the
pseudoscalar decay constant:
\begin{equation}
  \label{eq:fpsfss}
  L F_\mathrm{PS} = \mathcal F(L^{y_m} m)\, .
\end{equation}
Finite size scaling curves are displayed in Fig.~\ref{fig:su2fss},
where different values of the scaling exponent $y_m=1+\gamma_*$ are
used for rescaling the data. The plot indicates that the numerical
results are consistent with the existence of an IRFP with a low value
for $y_m$.

\begin{figure}[ht]
  \centering
  \includegraphics[height=2.0truein]{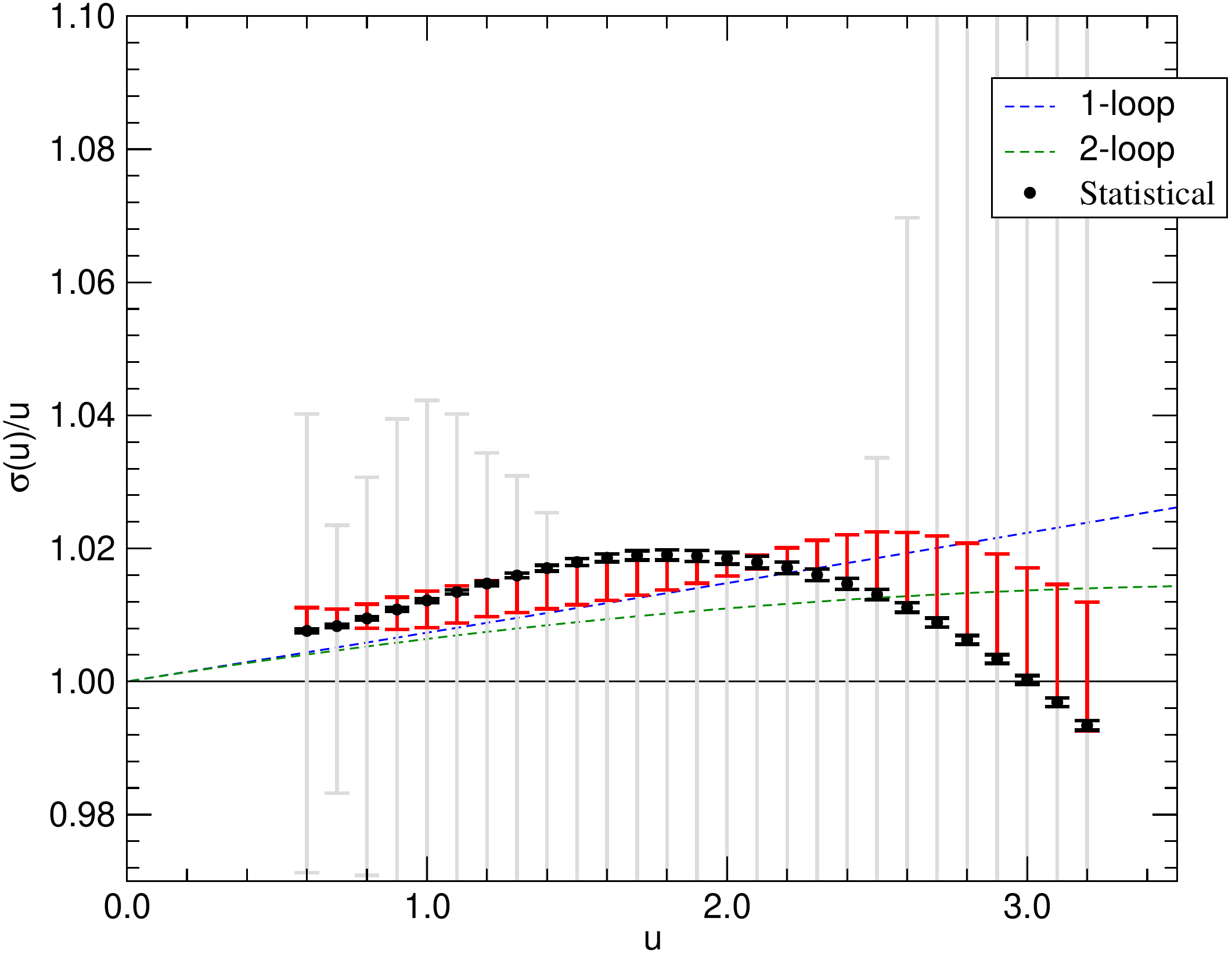}
  \includegraphics[height=2.0truein]{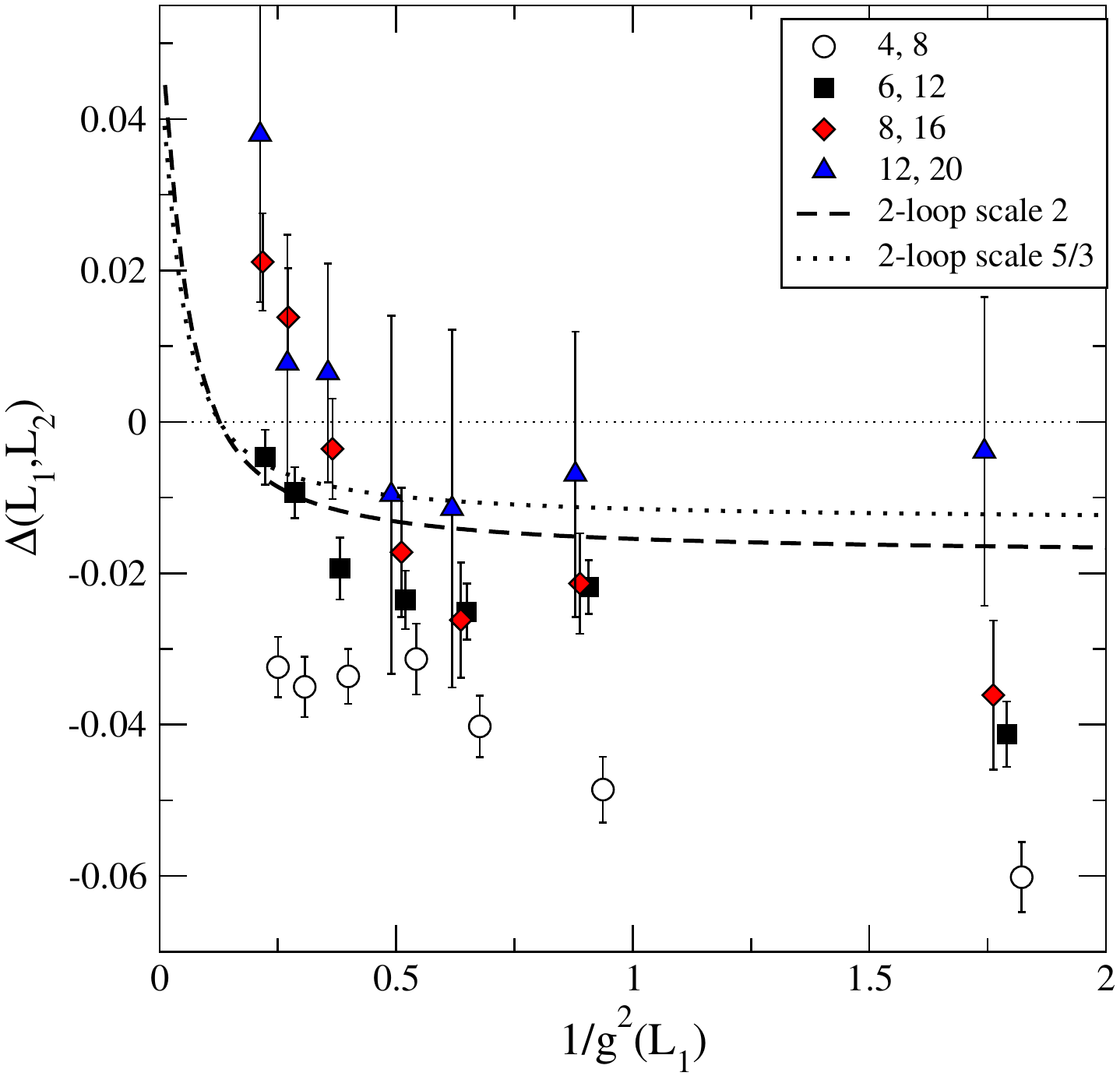}
  \caption{Running of the coupling constant in the SF scheme. Data are
    taken from Ref.~\cite{Bursa:2009we} for the left plot, and
    Ref.~\cite{Hietanen:2009az} for the right one. The large error
    bands in the plot on the left are due to the continuum
    extrapolation of the step scaling function.}
  \label{fig:su2sf}
\end{figure}

Schr\"odinger functional studies for this theory show some evidence in
favour of the existence of a conformal fixed point. The running of the
gauge coupling was studied in
Refs.~\cite{Hietanen:2009az,Bursa:2009we}. The results are summarized
in Fig.~\ref{fig:su2sf}. The plot on the left shows the continuum step
scaling function, an IRFP is signalled by the condition
$\sigma(u)/u=1$. The plot on the right is simply the difference of the
renormalized couplings at scales $L_1$ and $L_2$; with the latter
choice of variables the IRFP is identified by the difference of the
two couplings, $\Delta(L_1,L_2)$, changing sign. Note that in this
second way of presenting the data, the continuum limit is not taken,
but the lattices with larger values of $L_1$ are closer to the
continuum limit. In both cases, we see that the data are compatible
with the existence of a fixed point. However the current error on the
data (especially when performing the continuum extrapolation) is too
large to locate precisely the critical value $g^*$. This is not
surprising if the theory is really conformal, or near the edge of the
conformal window. When the running of the coupling becomes very slow,
the SF simulations have to resolve a very small signal that is easily
obscured by the statistical and systematic errors. This is a common
problem of {\em all}\ the studies of the running coupling for
conformal theories.

\begin{figure}[ht]
  \centering
  \includegraphics[width=3.0truein]{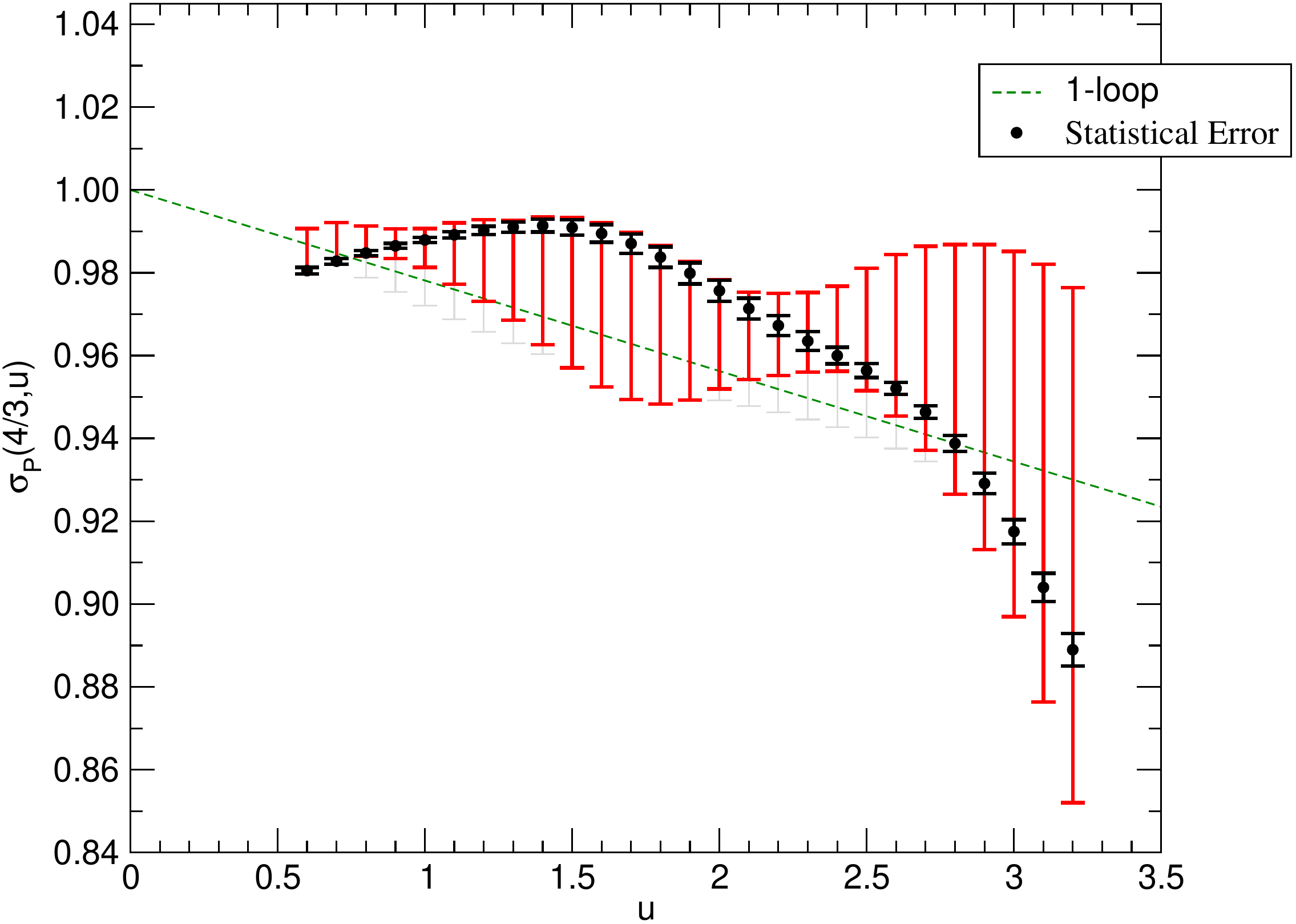}
  \caption{Running of the mass computed from the
    SF~\cite{Bursa:2009we}.}
  \label{fig:su2andim}
\end{figure}

Finally, as discussed in the previous Section, the SF can be used to
study the running of the mass and therefore deduce the anomalous
dimension $\gamma$~\cite{Bursa:2009we}.  Data for $\sigma_P(u)$ are
displayed in Fig.~\ref{fig:su2andim}.
Note that the step scaling function $\sigma_P$ is not small for a
conformal theory, and therefore can be measured with better
accuracy. Interestingly its value does not display any statistically
significant deviation from the one-loop prediction.

The anomalous dimension $\gamma_*$ can be read from the curve in
Fig.~\ref{fig:su2andim}, as long as the value of the gauge coupling at
the fixed point is precisely known. Unfortunately, as discussed above,
the latter cannot be located precisely; the uncertainty on the value
of $g^*$ is the main source of uncertainty in the determination of the
mass anomalous dimension. With the current data the best estimate is:
\begin{equation}
  \label{eq:su2gamma}
  0.05 < \gamma_* < 0.56\, .
\end{equation}
Preliminary results using the MCRG technique seem to confirm this
estimate~\cite{Keegan:lat10}.  

Recent results confirming the picture above have recently
appeared~\cite{DeGrand:2011qd}.

Results obtained with different techniques are consistent with the
existence of an IRFP for these theory, with a small value of the
anomalous dimension. The phenomenological consequences of such a small
value need to be investigated carefully.

\subsection{SU(3) with sextet fermions}
\label{sec:su32s}

Results have also been obtained for the SU(3) gauge theory with
$n_f=2$ in the two-index symmetric representation, i.e.\ the sextet
representation of SU(3). This model has also been proposed for
phenomenological applications under the name of Next to Minimal
Walking Technicolor (NMWT).

Results for the spectrum and the eigenvalues of the Dirac operator
have been presented in
Refs.~\cite{DeGrand:2008dh,DeGrand:2008kx,Fodor:2008hm,Fodor:2009ar,Fodor:2009nh,DeGrand:2009et,DeGrand:2009hu}.

\begin{figure}[ht]
  \centering
  \includegraphics[width=0.4\textwidth]{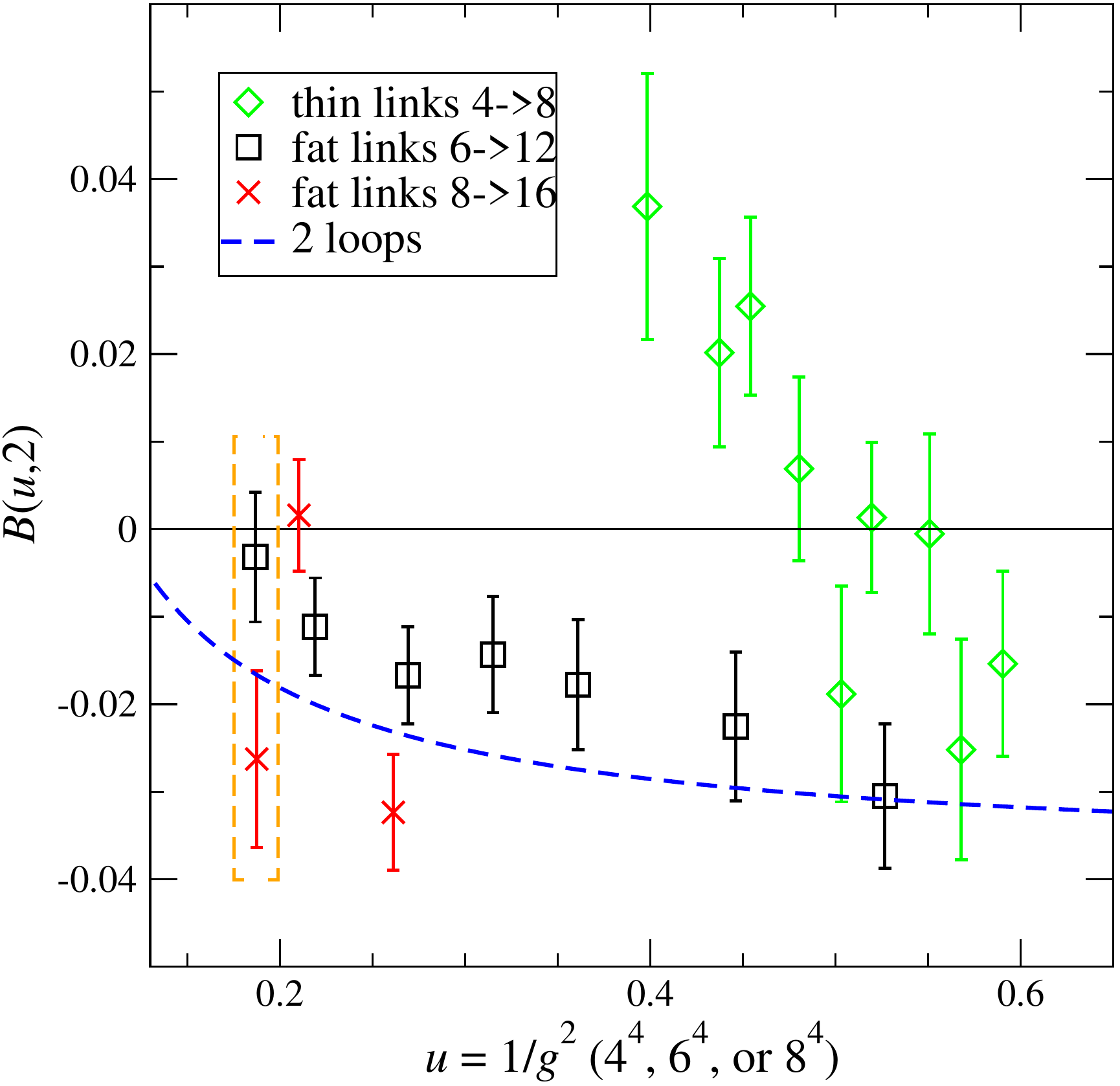}
  \includegraphics[width=0.39\textwidth]{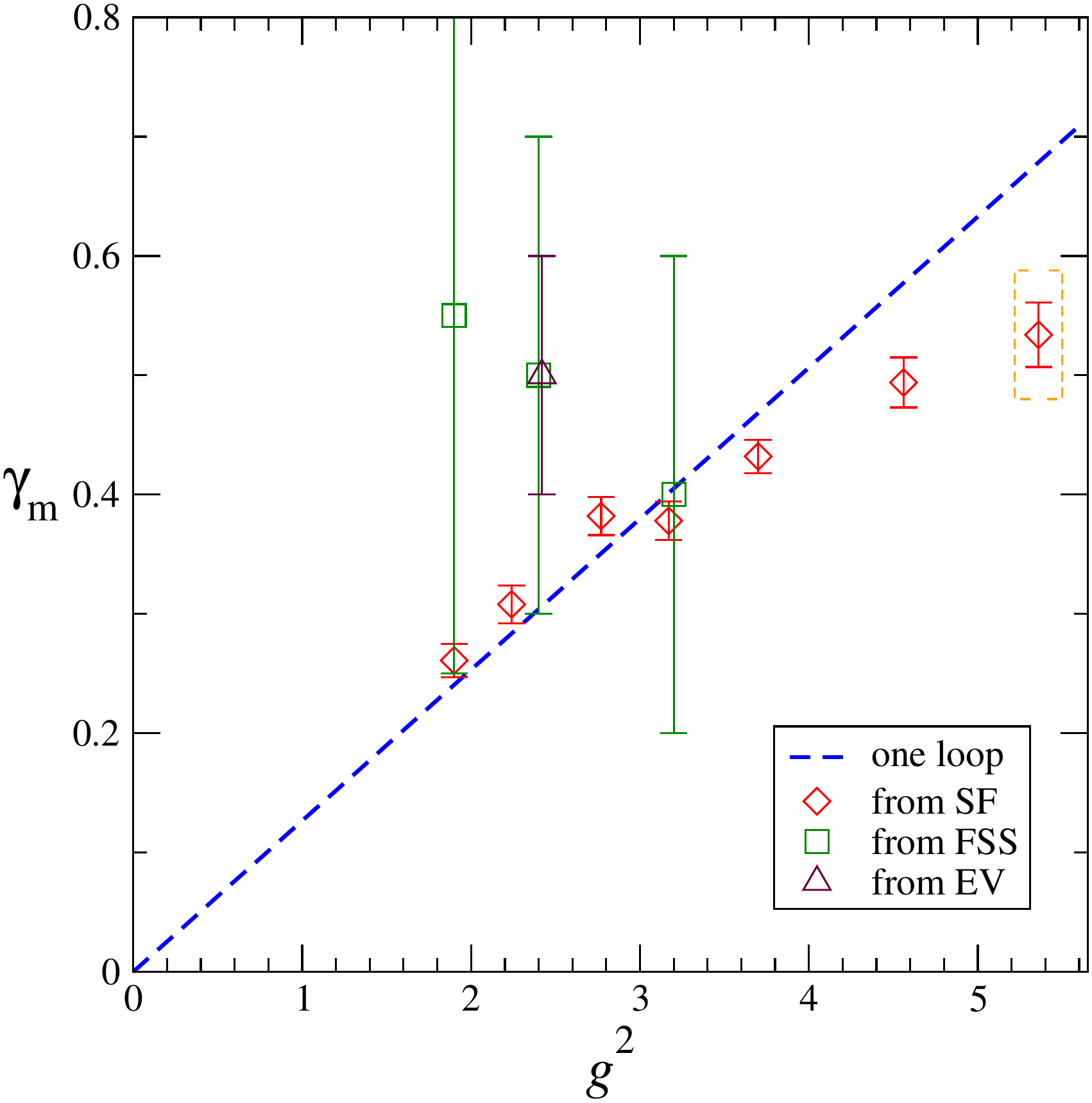}
  \caption{Running of the coupling (left) and the mass (right) for the
    SU(3) theory with two flavors in the two-index symmetric
    representation~\cite{Svetitski:lat10}.}
  \label{fig:b2}
\end{figure}

Results for the SF lattice step scaling function $\Sigma(u)$ have been
presented at this Conference~\cite{Svetitski:lat10}, showing evidence
of a slow running of the coupling. Different discretizations yield
statistically inconsistent estimates for the location of the fixed
point coupling $g^*$, emphasizing the importance of controlling the
size of lattice artefacts in these studies. The difference
\begin{equation}
  \label{eq:Bu2}
  B(u,2) = -1/g^2(L) + 1/g^2(2L)
\end{equation}
is displayed on the left-hand side in Fig.~\ref{fig:b2}, computed
using thin links ($\diamond$) and fat links ($\square$)
respectively. The difference between the diamond curve and the square
curve shows the effect of lattice artefacts.

Once again, results for the mass anomalous dimension have a much
smaller relative error, as seen in the plot on the right of
Fig.~\ref{fig:b2}. The estimates obtained from the SF, from
finite size scaling, and from the eigenvalues of the Dirac operator
are in broad agreement, and suggest the bound $\gamma_* < 0.6$.

Studies at finite temperature on the other hand indicate that the
theory is outside the conformal window and ``slow
walking''~\cite{Kogut:2010cz,Sinclair:lat10}.

A detailed computation of the spectrum has been performed with
staggered fermions for the same theory, and presented at this
Conference~\cite{Kuti:lat10}. The numerical data in this case are
consistent with chiral symmetry breaking.

\begin{figure}[ht]
  \centering
  \includegraphics[width=0.8\textwidth]{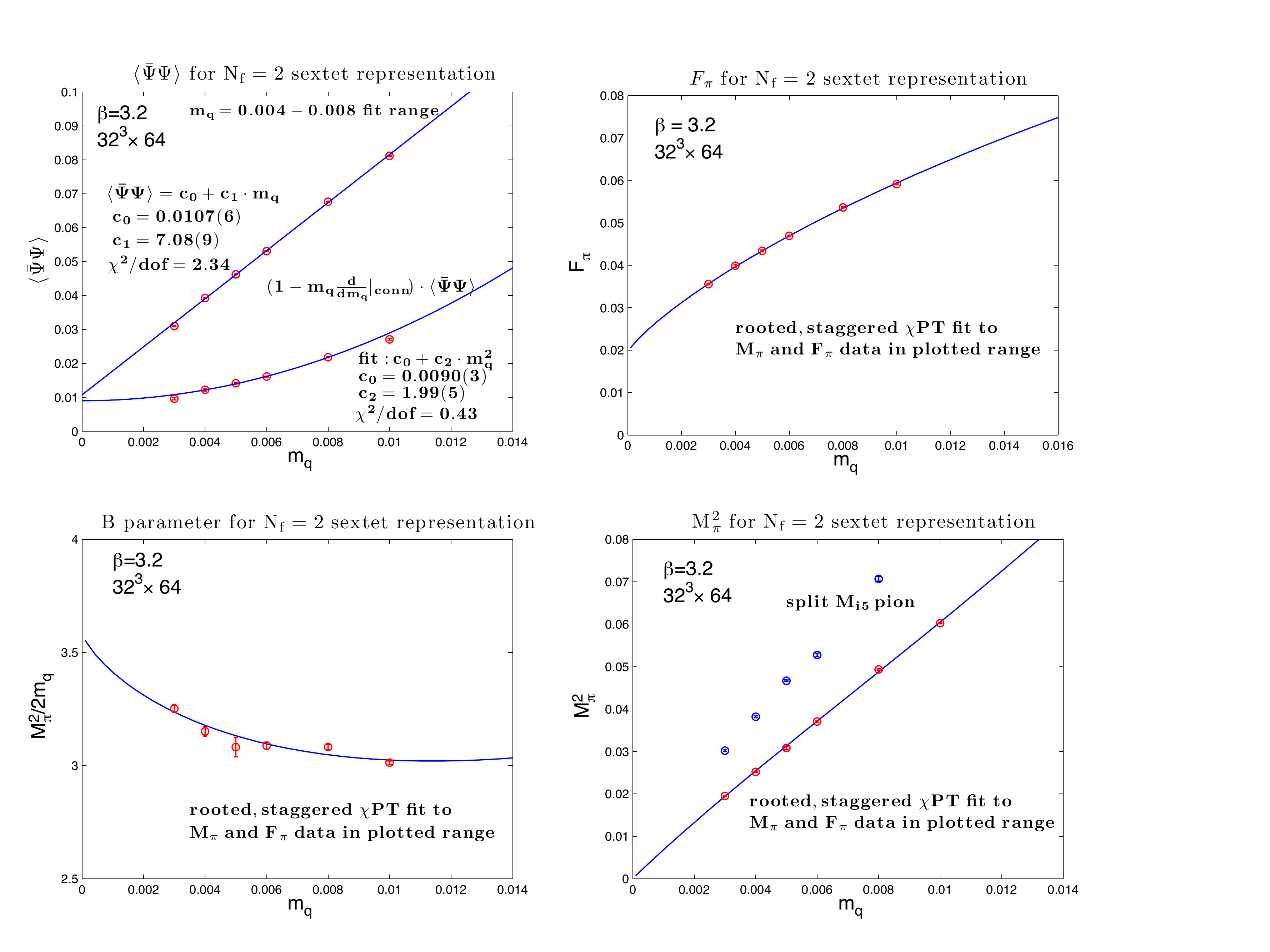}
  \caption{Chiral extrapolation of the chiral condensate and other
    spectral quantities. Data are consistent with rooted staggered
    chiral perturbation theory~\cite{Kuti:lat10}. }
  \label{fig:k2sext}
\end{figure}

Further work is needed to reach consensus on the long-distance
dynamics in this case.

\section{Outlook}
\label{sec:out}

There have been numerous studies of nonperturbative dynamics beyond
QCD in the last two years. Several theories that are expected to be at
the edge of the conformal window have been investigated, using the
tools described in Section 2. This is an interesting problem in field
theory, and there are several intriguing data, suggesting that first
evidence for the existence of fixed points has been found. However
robust results can be obtained only by keeping the systematic errors
under control.

In particular simulations must be performed at small masses, on large
volumes, trying to minimize lattice artefacts that can obscure the
physically interesting features. Preliminary results about simulations
with improved actions for the SU(2) theory have been presented in
Refs.~\cite{Karavirta:2010ef,Mykkanen:2010ym}. At the same time it is
worthwhile to keep looking for better observables that can yield
unambiguous signals.

It is important to develop a strong link with the phenomenological
work and the data analyses at the LHC. Lattice data will have an
impact on phenomenology only if the relevant questions are identified
and answered in a quantitative way. A wishlist of interesting issues
has been discussed by Chivukula at this
Conference~\cite{Chivukula:2010xz}.

\acknowledgments It is a pleasure to thank my collaborators, F~Bursa,
S~Catterall, M~Frandsen, J~Giedt, B~Lucini, A~Patella, C~Pica,
T~Pickup, A~Rago, F~Sannino, R~Zwicky, who have contributed to many of
the results presented in this review. I would also like to thank
R~Brower, S~Chivukula, T~DeGrand, G~Fleming, A~Hasenfratz, K~Holland,
J~Kuti, D~Lin, MP~Lombardo, D~Nogradi, T~Onogi, E~Pallante, E~Simmons,
B~Svetitsky for many illuminating discussions and for providing me
with results ahead of (and during) this conference. I am indebted to
the Aspen Center for Physics, and the Theory Unit
at CERN, for organizing fruitful workshops where these topics have
been actively discussed. LDD was supported at the time of the
conference by an STFC Advanced Fellowship. The work presented here has
also been supported by a travel grant from the Royal Society of
Edinburgh.

\bibliographystyle{unsrt}

\end{document}